\documentclass[a4paper,12pt]{article}
\usepackage{amsmath,amssymb,amsthm,bm,empheq,extarrows, array, graphicx, color, hyperref,placeins}
\usepackage[paper=letterpaper,margin=1.0in]{geometry}
\graphicspath{{./figures/}}

\newcommand{\be}{\begin{equation}}
\newcommand{\ee}{\end{equation}}
\newcommand{\bea}{\begin{eqnarray}\displaystyle}
\newcommand{\eea}{\end{eqnarray}}
\newcommand{\bi}{\begin{itemize}}
\newcommand{\ei}{\end{itemize}}
\newcommand{\bfig}{\begin{figure}}
\newcommand{\efig}{\end{figure}}
\newcommand{\nn}{\nonumber}
\newcommand{\eref}[1]{(\ref{#1})}
\newcommand{\boxeq}[1]{\begin{empheq}[box={\fboxsep=10pt\fbox}]{align} #1 \end{empheq}}
\newcommand{\gau}[2]{\left[#1\right]#2}

\newcommand{\la}{\langle}
\newcommand{\ra}{\rangle}

\def\Im{ \mathrm{Im} }

\def\s{\sigma}

\def\rd{\mathrm{d}}

  \def\cC{{\cal C}}
  
\def\cG{{\cal G}}  \def\cI{{\cal I}}
  
\def\cM{{\cal M}}  
  
 \def\cT{{\cal T}}

\def\tr{ {\rm tr } } 
\newcommand{\cor}[1]{\langle{#1}\rangle}

\begin{document}

\rightline{QMUL-PH-14-21}

\vspace*{2cm} 

{\LARGE{  \bf 
\centerline{Permutation combinatorics of } }
{ \LARGE \bf  \centerline {worldsheet moduli space} } }

\vskip.5cm 

\thispagestyle{empty} \centerline{
    {\large \bf Laurent Freidel,
${}^{a,} $\footnote{ {\tt lfreidel@perimeterinstitute.ca}}}
    {\large \bf David Garner,
${}^{b,} $\footnote{ {\tt d.p.r.garner@qmul.ac.uk}}}
   {\large \bf and Sanjaye Ramgoolam
               ${}^{b,}$\footnote{ {\tt s.ramgoolam@qmul.ac.uk}}   }
}              
               
\vspace{.4cm}

\centerline{{\it ${}^a$ Perimeter Institute for Theoretical Physics,}}
\centerline{{\it 31 Caroline Street North, Waterloo, Ontario N2L 2Y5, Canada}}

\vspace{.4cm}

\centerline{{\it ${}^b$ Centre for Research in String Theory,}}
\centerline{ {\it School of Physics and Astronomy},}
\centerline{{ \it Queen Mary University of London},} 
\centerline{{\it    Mile End Road, London E1 4NS, UK}}

\vspace{1.4truecm}

\thispagestyle{empty}

\centerline{\bf ABSTRACT}
\vspace*{.5cm} 

Light-cone string diagrams have been used to reproduce the orbifold Euler characteristic of moduli spaces of 
punctured Riemann surfaces at low genus and with few punctures. Nakamura studied 
the meromorphic differential introduced by Giddings and Wolpert to characterise light-cone diagrams and
introduced a class of graphs related to this  differential. These  Nakamura graphs were used to
parametrise the cells in a light-cone cell decomposition of moduli space. We develop links between Nakamura graphs and realisations of the worldsheet as branched covers. This leads to  a development of  the combinatorics of Nakamura graphs in terms of permutation tuples.  For certain classes of cells, including those of top dimension, there is a simple relation to Belyi maps, which allows us to use results from Hermitian and complex matrix models to give analytic formulae for the counting of cells at arbitrarily high genus. For  the most general cells, we develop a new equivalence relation on Hurwitz classes which organises the cells and allows efficient enumeration of Nakamura graphs using the group theory software GAP. 

\vskip.4cm

\setcounter{page}{0}
\setcounter{tocdepth}{2}

\newpage

\tableofcontents

\section{Introduction } 

The light-cone gauge in string theory involves only physical degrees of freedom and 
leads to a manifestly unitary $S$-matrix, while Lorentz invariance appears non-trivially \cite{mandelstam73,mandelstam85}. 
The computation of string amplitudes uses light-cone diagrams, parametrised 
by string length and twist parameters along with interaction times, where the lengths of the strings are proportional to
the light-cone momenta. The covariant gauge has manifest Lorentz invariance 
but unitarity is non-trivial. String amplitudes are calculated by integration over the moduli space of Riemann surfaces $\cM_{ g , n } $, for surfaces of genus $g$ and $n$ punctures.  

In the paper \cite{GiddWol}, Giddings and Wolpert showed that each closed string light-cone diagram determines a worldsheet 
equipped with a meromorphic one-form with purely imaginary periods and residues that sum up to zero.
The meromorphic one-form (or {\it Giddings-Wolpert differential}) was constructed by disassembling the light-cone diagram into a
number of strips on each of which the meromorphic one-form is trivial, and then making identifications on the boundaries of the
strips. It was explained there that light-cone string diagrams lead to a single cover of moduli space, which is important for an equivalence of the light-cone formulation to the covariant formulation. 

In the paper \cite{Nak}, Nakamura developed the work of  \cite{GiddWol} and showed how to 
compute the orbifold Euler characteristic of $\cM_{g,n}$ using the cell decomposition coming from light-cone diagrams. The key  step was the introduction of  graphs, embedded on 
the worldsheet,  whose vertices are the zeroes and poles of the GW differential, and whose real trajectories form the edges of the graph.
The embedded graph (or ribbon graph)  inherits a cyclic order at the vertices - a familiar property which also arises in large $N$ expansions of gauge theories.
Each such graph - which we call a {\it Nakamura graph} - corresponds to a cell in the space of GW differentials on a surface of genus $g$ with $n$ punctures.
These cells are quotiented by the symmetry group of the graph to obtain cells in $\cM_{g,n}$. 
The dimension of each cell in this {\it light-cone cell decomposition} can easily be read off from the
structure of the Nakamura graph. 
The graphs were counted for low values of $g$ and $n$, and the dimensions and symmetries of the graphs were used to calculate the orbifold Euler characteristic of the moduli space $\cM_{g,n}$.
These results agreed with the result for general $g$ and $n$ computed by Harer and Zagier \cite{HZ}.

There is, as yet, no proof that the light-cone approach reproduces the 
orbifold Euler characteristic in general. However the evidence that this is true is highly non-trivial: a large number of graphs were counted to verify this in \cite{Nak}. 
This is a very important result, since it implies that  the Nakamura graphs contain 
all the information needed to describe precisely how light-cone diagrams can be used to give a
single cover of moduli space. This approach implicitly resolves the technical issue \cite{GiddWol,dhokPhong}  of giving a precise specification of the region in the space of light-cone (LC) string parameters which covers every point 
in $\cM_{g,n}$ precisely once. A naive integration without restrictions would lead to an overcounting problem discussed in  \cite{GiddWol} and, as anticipated there, its solution should involve systematics similar to those encountered in  Feynman graphs. The work of \cite{Nak} associates a  cell  of moduli space to each Nakamura graph.  The use of graphs in the LC cell decomposition is analogous to the graphs in the Kontsevich-Penner (KP) cell decomposition
of decorated moduli space \cite{Penner,Kontsevich}. Indeed, the KP cell decomposition 
has been used to compute homology groups  and  intersection numbers of Mumford-Morita classes on moduli space. 
The LC and KP cell decompositions both involve graphs with cyclic orientation at the vertices (ribbon graphs). 
However, the Nakamura graphs are much more restricted because of certain causality relations controlling 
the connectivity of the vertices.  As a result the LC cell decomposition requires fewer cells, and so is much more
economical \cite{Nak}.

Moreover the Giddings-Wolpert differential and the Nakamura decomposition of Riemann surfaces into strips is an essential ingredient 
of the newly formulated metastring \cite{meta}. Since the metastring is chiral, it is necessary for its formulation to provide a
parametrisation of the moduli space of Riemann surfaces that includes a notion of worldsheet time while preserving modular invariance. The Nakamura graphs and their implied strip decomposition do exactly this.

A detailed understanding of the topology of $ \cM_{ g , n  }$ is fundamental both to mathematics 
and string theory. The KP cell decomposition is well studied in mathematics and has also been used recently 
in describing the link between string theory integrals and Feynman integrals \cite{Tourkine}.  
In another line of development, the systematics of a variety of  Feynman graph counting problems 
of quantum field theory and  ribbon graphs of large $N$ matrix theories have found a unifying description in terms of permutations in \cite{dMKRam2010,dRW1209}, with group theoretic structures such as double cosets
playing a central role.  The present paper initiates a systematic study of Nakamura's cell decomposition. 
We develop a general description of  the combinatorics
of Nakamura graphs in terms of tuples (finite sequences) of permutations. 
We present three descriptions of the graphs in terms of permutations in this paper. Two of them involve triples 
of permutations, and are closely related to the known fact that ribbon graphs can be described in terms of triples 
\cite{Grothendieck,schneps}.
Since a Nakamura graph is not a generic ribbon graph, but rather a ribbon graph subject to non-trivial causality conditions, 
the associated permutation triples satisfy some non-trivial constraints.
The third description of a Nakamura graph involves a tuple of up to $(l+2)$ permutations, where $l$ is the number
of interaction vertices in the light-cone diagram.
This description requires more permutations in general to describe the graph than in the other two descriptions,
but has the advantages that the permutations live in a permutation group of smaller degree, and also that the 
causality conditions are much simpler. The permutations in this description are elements of $S_d$, where $d$ 
is the number of faces of a Nakamura graph, or equivalently the number of edges of the graph connecting to 
poles of the GW differential with positive residues. We call this description the $S_d$ description.

For Nakamura graphs corresponding to the top-dimensional cells of moduli space, the $S_d$ description
can be simplified further. In this case, $d$ has to be even and the tuple has exactly three permutations.
The counting of Nakamura graphs for these cells is a counting of permutation triples, where one of the permutations consists of 
$d/2$ cycles of length $2$. This permutation counting is exactly the one that arises in correlators
of the Hermitian matrix model, which have been related to branched covers of the sphere \cite{KRBelyi1002,IdF,IB}.
This allows us to draw upon exact results on generating functions for Matrix model 
correlators \cite{HZ} to give analytic expressions for the contribution to the Euler characteristic from 
the top-dimensional cells, for any $g$ and $n$. The combinatorics of non-zero codimension cells  
is more non-trivial. A precise permutation description is nevertheless possible. 
We expect it to lead to analytic results in the future. For  the current paper, 
we have developed a computer algorithm based on this description, which reproduces 
all the tables from Nakamura and extends them to higher $g$ and $n$. 

We now describe the content of the paper in more detail. 
In Section \ref{sec:review}, we start by recalling the properties of the Giddings-Wolpert differential  \cite{GiddWol} 
and explaining how Nakamura associated a graph to each differential \cite{Nak}. 
The parameters describing the cells in light-cone cell decompositions are introduced.
For fixed $g$ and $n$, the integer $d$ gives the total number of edges incident on the poles of the GW differential with positive residue (which we call {\it incoming poles}).
It is also the number of strips which can be glued together to produce the worldsheet;
each strip is incident on one incoming pole and one outgoing pole. 
The {\it branching constant} $\Delta$ is an integer describing the combined orders of all the zeroes and their departure from simplicity; when all the zeroes are simple, then $\Delta=0$.
The number of internal edges is denoted by $I$; these are the edges of the Nakamura graph which connect zeroes of the GW
differential directly to zeroes.
The top dimensional cells of the LC cell decomposition only involve simple zeroes of the GW differential and 
their associated Nakamura graphs have no internal lines, so $\Delta=I=0$ for cells at top dimension. Lower 
dimensional cells can involve higher order zeroes as well as real trajectories connecting the zeroes.

In Section \ref{sec:belyinak} we relate Nakamura graphs to dessins d'enfants and Belyi maps.
A dessin d'enfant is a bipartite graph embedded on a surface with a cyclic ordering of the edges at each vertex.
Bipartite graphs have two types of vertices, which can be coloured in black or white, in which each edge connects to two
vertices of different colours.
We can convert a Nakamura graph to a dessin by introducing auxiliary vertices along the edges of the graph in such a way that
the graph becomes bipartite. The structure of these graphs can then be described by a triple of permutations, which
also allow the graphs to be related to branched covers of the Riemann sphere known as {\it Belyi maps} \cite{Belyi,Grothendieck,schneps}. 
The simplest way to convert a Nakamura graph to a bipartite graph is to subdivide every edge; this graph has $4d+2I$ edges,
so can be described by a triple of permutations in the symmetric group $S_{4d+2I}$.
There is also another general way to convert a Nakamura graph into a bipartite graph which requires fewer subdivisions
of edges, which allows a description in terms of a triple of $S_{2d+2I}$ permutations.
While every Nakamura graph has a description in terms of these triples of permutations, not all permutation triples give
Nakamura graphs; in Section \ref{sec:triplestocells}, we state the required properties that a permutation triple
must satisfy to give a Nakamura graph.

In Section \ref{sec:hurwitz}, we develop a new permutation description of Nakamura graphs by considering branched covering maps from the   worldsheet onto the 
infinite cylinder - equivalently, by composing with a conformal map, branched covering maps to  the Riemann sphere. 
 The section starts with a review of branched covers, and their 
description in terms of equivalence classes of tuples of permutations which we call {\it Hurwitz classes}. 
The branched covers can be constructed by a gluing construction on the $d$ faces (strips) of the Nakamura graph.
The degree of the branched cover of the sphere associated to a graph is $d$. The branch points of the 
cover are related to the vertices of the Nakamura graph.
Each Hurwitz class determines a unique Nakamura graph, but there can be multiple Hurwitz classes corresponding
to a given Nakamura graph. To solve this redundancy, we introduce an equivalence relation on the space of Hurwitz
classes which we call {\it slide-equivalence}. This equivalence relation is related to  the fact that the connectivity of a Nakamura graph does not  determine the relative time-ordering of the zeroes (interaction vertices) of the GW differential.
There is a one-to-one correspondence between slide-equivalence classes and Nakamura graphs.

In Section \ref{sec:hmm} we explore some links between the counting of cells in the moduli space and the correlators of matrix models. 
Cells of top dimension in the LC decomposition are specified by Nakamura graphs with simple zeroes and no internal edges.
Within the slide-equivalence class of a top-dimensional graph, there is a unique Hurwitz class consisting of a tuple of three permutations. This permutation triple naturally corresponds to a Belyi map (a covering of the sphere branched at three points), without the need to introduce new vertices or subdivide edges of the Nakamura graph.
The counting of Belyi maps is known to be related to correlators of the Hermitian matrix model \cite{KRBelyi1002,IdF,IB}. 
This allows us to use known exact results from Hermitian matrix models \cite{HZ} to obtain all orders 
analytic formulae for the contribution to the Euler characteristic from the top-dimensional cells of the LC cell decomposition. 
These results agree with the tables given by Nakamura for small $g$ and $n$.
We can also consider cells with lower dimension with branching constant $\Delta>0$ and no internal edges ($I=0$).
In this case, we can use complex matrix models to derive analytic formulae for the contributions to the Euler characteristic from lower-dimension cells. (At the present stage, we have no map to matrix models for the counting of the most general cells involving $I > 0$.)

Finally, in Section \ref{sec:gap}, we test computationally the validity of the LC cell decomposition and its description in terms of slide-equivalences of Hurwitz classes.
Using the group theory software GAP \cite{GAP}, we use the $S_d$ description to enumerate the cells
and their dimensions in terms of Nakamura graphs, reproducing and extending the tables found in \cite{Nak}.
The computation is significantly facilitated by the introduction of the concept of an {\it $I$-structure}, which 
contains some coarse information about the internal edges of a Nakamura graph. 
It is an invariant of the slide-equivalence classes of Hurwitz-classes.  Double cosets of $S_d$ also play a role  
in the computation. We conclude with some discussion of our results and possible future directions.

\section{Review: Giddings, Wolpert and Nakamura} \label{sec:review}
\subsection{The Giddings-Wolpert differential}
Let $\Sigma$ be a Riemann surface with $n$ marked points $P_1, P_2, \ldots, P_n$ and genus $g$, where $n\geq 2$.
Associate a set of real numbers $r_1, r_2, \ldots r_n$ respectively to the $n$ marked points, which satisfy $\sum_i r_i=0$.
Giddings and Wolpert proved in \cite{GiddWol} that there exists a unique abelian differential $\omega$
on the Riemann surface $\Sigma$ such that $\omega$ has $n$ simple poles at the points $P_i$ with respective residues $r_i$,
and pure imaginary periods on any closed integral on the surface.

The Giddings-Wolpert differential $\omega$ yields a global time coordinate on the surface, up to an overall constant representing the time translation symmetry. If we fix a point $z_0$ on the surface which is not a pole of $\omega$, 
then we can define the global time coordinate of a generic point $z$ on the surface to be $T := \mathrm{Re}(\int_{z_0}^z \omega)$.
This expression does not depend on the choice of integration contour from $z_0$ to $z$, since any two paths from 
$z_0$ to $z$ differ only by a closed contour, and the integral of the differential along any closed contour is imaginary.
The global time coordinate tends to positive infinity as we approach the poles with negative residues, and to
negative infinity as we approach the poles with positive residue. 
We call the poles with positive residue the {\bf incoming poles}, and the poles
with negative residue the {\bf outgoing poles}.

For the cases of the sphere and the torus, it is straightforward to construct the GW differential of a given marked surface
and its time coordinate explicitly. Take a sphere with $n$ marked points $P_i$ and associated reals $r_i$, where $\sum_i r_i=0$. We can choose coordinates $z$ on the sphere such that the marked points $P_i$ are located at $z=p_i$ for some $p_i\in\mathbb{C}$. In this chart, the GW differential can be explicitly written as
\bea
\omega(z;p_i) := \sum_{i=1}^n \frac{r_i\rd z}{z-p_i}.
\eea
It is clear that this differential has residues $r_i$ at the points $P_i$, and that the integral of the differential along any closed contour $\mathcal{C}$ is $\oint_\mathcal{C}\omega = 2\pi i \sum_{P_i \in \mathcal{C}}r_i$, which is purely imaginary.
The global time coordinate is
\bea
T(z)= \ln \left( \prod_i |z-p_i|^{r_i} \right) + T_0,
\eea
where $T_0$ is an arbitrary constant.

Now consider a torus with $n$ marked points $P_i$, associated real values $r_i$ with $\sum_i r_i=0$, and
modular parameter $\tau$ with $\Im(\tau)>0$.
This torus can be realised as the quotient of the complex plane $\mathbb{C}$ by the equivalence relation $z \sim z+n+m\tau$, where $n$ and $m$ are integers.
In these coordinates, the marked points $P_i$ are located respectively at $z=p_i$ for some $p_i=a_i+b_i\tau$, 
where $0\leq a_i,b_i<1$.
To define the GW differential on this surface, we introduce the Jacobi theta function $\theta_{11}(z;\tau)$, which is a holomorphic quasi-periodic function on the complex $z$ plane satisfying
\bea
\theta_{11}(z+1; \tau) =\theta_{11}(z; \tau),&  \qquad   &\theta_{11}(z+\tau; \tau) = e^{-2\pi i(z+1/2)}\theta_{11}(z;\tau), \\
\theta_{11}(z;\tau+1) = \sqrt{i}\theta(z; \tau),&  \qquad  &\theta_{11}(z/\tau; -1/\tau) = (-i)\sqrt{i\tau}e^{i\pi z^2/\tau}\theta_{11}(z;\tau), \label{eq:thetaprops}
\eea
and behaves like $\theta_{11}(z;\tau)\approx z$ for small values of $z$.
The GW differential on this surface is 
\bea
\omega(z;p_i, \tau) := dz\sum_{i=1}^nr_i\left(-2\pi i \frac{\Im(p_i)}{\Im(\tau)}+\frac{\theta'_{11}(z-p_i;\tau)}{\theta_{11}(z-p_i;\tau)} \right),
\eea
and the associated global time coordinate on the surface is 
\bea
T(z) = \sum_i r_i\left[ 2\pi \frac{\Im(p_i)}{\Im(\tau)}\Im(z) + \log |\theta_{11}(z-p_i;\tau)|\right] + T_0,
\eea
where $T_0$ is an arbitrary constant.
It can be shown from the above properties and relations of the Jacobi theta function that $\omega(z; p_i)$ and $T(z)$ are well-defined on the torus, i.e. these definitions are invariant under the coordinate shifts $z\to z+m+n\tau$ and under the modular transformations $(\tau, p_i)\to (\tau+1,p_i)$, $(\tau, p_i) \to (-1/\tau, p_i/\tau)$.
It can also be seen that the integrals of the differential along the cycles $a: z\to z+1$ and $b: z\to z+\tau$ are imaginary,
and that each pole $p_i$ has residue $r_i$, so that all the periods are pure imaginary. 
Formulae for Giddings-Wolpert differentials in terms of theta functions at genus one and higher can be found in recent work 
\cite{ishmur1307}. 

\subsection{Nakamura graphs} \label{sec:props}

The Giddings-Wolpert differential associated to a marked Riemann surface naturally gives rise to an embedded 
ribbon graph on the surface. This construction was developed by Nakamura in \cite{Nak}, and leads to a cell decomposition
of the moduli space of Riemann surfaces in which each cell is specified by a graph. In this section we review the basic properties 
of these graphs, which we call {\it Nakamura graphs}.

Consider a marked Riemann surface $\Sigma$ with GW differential $\omega$. The GW differential has poles at the $n$ marked points $P_i$ with residues $r_i$. 
For any unmarked point on $\Sigma$, we can choose local complex coordinates $z$ around that point such that $\omega= d(z^{m+1})$ for some $m$. A zero of order $m$ of the GW differential is a point at which $m>0$.
For each point on the surface, there exists a set of directions in which $z^{m+1}$ is real  - these are the {\it real trajectories}
that extend out from the point.
A zero of order $m$ has $2(m+1)$ real trajectories extending out from the zero. If $m=1$, the zero is called {\it simple}.  
Real trajectories extending out from the zeroes of the GW differential will only meet at poles and zeroes of the differential.

\begin{figure}[t]
\begin{center}
\includegraphics[width=0.35\textwidth]{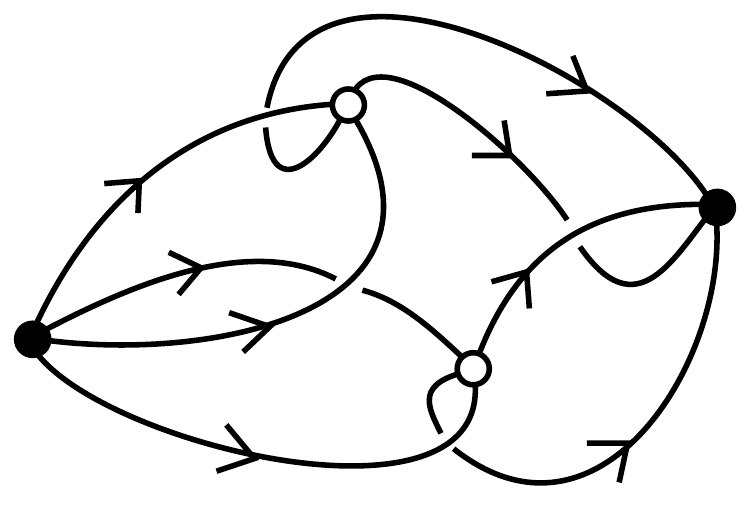} \qquad
\includegraphics[width=0.4\textwidth]{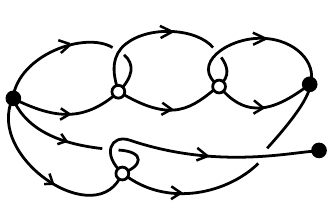}
\caption{Two examples of Nakamura graphs.}
\label{fig:unlabelled}
\end{center}
\end{figure}

The set of real trajectories that extend out from all the zeroes of the GW differential define a ribbon graph embedded onto the surface, with the vertices of the graph corresponding to the poles and zeroes of $\omega$, and the edges of the graph corresponding to the real trajectories.
The edges also inherit an orientation from the GW differential: they are oriented in the direction along which the global time coordinate increases.
Some examples of Nakamura graphs are shown in Figure \ref{fig:unlabelled}.

The Nakamura graph associated to a marked Riemann surface is uniquely determined by its Giddings-Wolpert differential.
It was shown by Nakamura in \cite{Nak} that such a graph always has the following properties:
\bi
\item The graph is connected, oriented, and cyclically ordered at the vertices.
\item The edges connecting to a pole are either all oriented towards the pole or all oriented away from the pole.
\item A zero connects to cyclically alternating incoming and outgoing edges, and has a valency of at least four.
\item No edge connects to the same end point twice, and no edge has only poles as its end points.
\item Every face of the ribbon graph contains on its boundary exactly two poles, one incoming and one outgoing.
\ei
Each face of the graph is bounded by two extended real trajectories of the GW differential. It is possible to choose local coordinates $z$ on each face such that $\omega=dz$ within the face, and where $z$ lies in the range
$0 < \Im(z) < b_i$ for some $b_i$. 
This means that each face of the graph is holomorphic to a strip $\mathbb{R}\times (0,b_i)$ in the complex plane,
and each strip has a width $b_i$ which is determined by the GW differential.
The combination of the Nakamura graph, the widths of the strips, the time coordinates of the zeroes, and the residues 
around the poles, is enough to reconstruct the Giddings-Wolpert differential on a surface, and hence to specify 
its complex structure.

\begin{figure}[t]
\begin{center}
\includegraphics[width=0.95\textwidth]{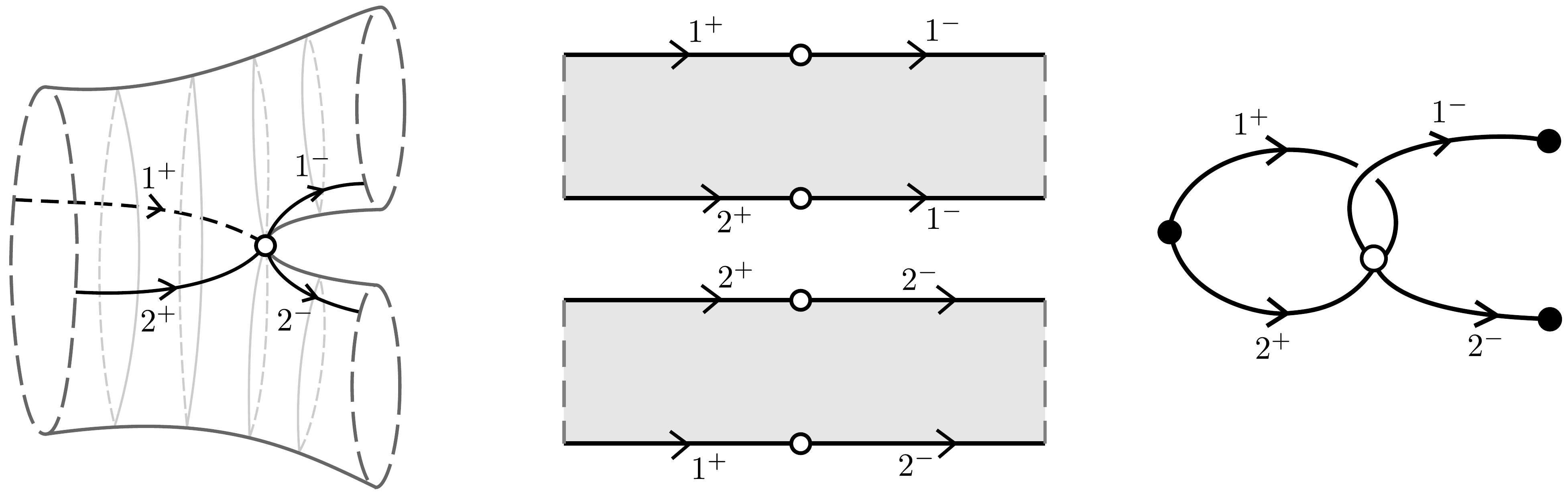}
\caption{A Riemann surface can be decomposed into glued strips via a Nakamura graph.}
\label{fig:pants}
\end{center}
\end{figure}

An example of the gluing of strips to give a surface with an embedded Nakamura graph is shown in Figure \ref{fig:pants}. A Riemann sphere with three punctures is conformally equivalent to a `pants' diagram, with the boundaries extended out to infinity. The GW differential on this surface traces out a Nakamura graph, given on the right of the figure, which partitions the pants diagram into two infinite strips. The poles of the GW differential are represented by black vertices of the Nakamura graph, and correspond to the boundaries of the strips located at positive and negative infinity. In this case, the widths of the strips are determined by the residues of the marked points.

If we take some Nakamura graph arising from a Giddings-Wolpert differential and consider all possible strip widths that are consistent with the specified residues at the poles, and all possible time coordinates of the zeroes that are consistent with the causal ordering of the zeroes, then we will in general find a family of inequivalent GW differentials that can arise from a single Nakamura graph. As each GW differential corresponds to a unique Riemann surface, this means that each Nakamura graph specifies a cell in $\cM_{g,n}$, the moduli space of inequivalent Riemann surfaces of genus $g$ with $n$ marked points.It was shown in \cite{Nak} that counting all such possible graphs can give information about the moduli space of Riemann surfaces. Nakamura successfully found all the graphs corresponding to surfaces with Euler characteristic $\chi := -(2g-2+n) \geq -6$, and used this to calculate the orbifold Euler characteristic of moduli space in many different cases.

\subsection{Parameters of Nakamura graphs and moduli space}\label{Nakpars}
We conclude this section by presenting some relevant relations between the parameters of   Nakamura graphs and their
associated cells in moduli space.

A Nakamura graph consists of $V$ vertices, $E$ edges, and $d$ faces.
The $V$ vertices are separated into $l$ zeroes and $n$ poles.
All edges connect to zeroes, and no edge connects two poles together.
There are exactly two poles on the boundary of each of the $d$ faces of the graph, one incoming and one outgoing.
Hence, there are $d$ external edges of the graph connecting incoming poles to zeroes, $d$ external edges connecting
outgoing poles to zeroes, and $I$ internal edges that connect only to zeroes. Summarising, we have
\bea
V&=& l+n, \nn \\
E&=& 2d+I, \nn \\
F&=& d.\nn
\eea
The Euler characteristic of a surface with an embedded graph is $2-2g=V-E+F$, which gives the relation
\bea
d+I-l = 2g-2+n.\label{init1}
\eea

Next, we consider the valencies of the vertices. As all faces have on their boundary exactly one incoming pole, 
the valencies of the incoming poles sum up to $d$, and similarly for the outgoing poles.
As the zeroes always border an equal number of incoming and outgoing edges, the valencies of the zeroes 
are always even. The zeroes correspond to the points where at least two real trajectories meet, and so the valency of a zero is always greater than four. We define the \emph{branching number} $\Delta$ to be
\bea
\Delta = \sum_{j=1}^l\left[ \left(\frac{v_j}{2}\right)-2\right], 
\eea
where the $v_j$ are the valencies of each of the $l$ zeroes.
The branching number is a non-negative integer for every Nakamura graph.
This sum rearranges to
\bea
2\Delta + 4l = \sum_{j=1}^l v_j. 
\eea
Now, adding the sum of the valencies of the poles to this equation give us the sum
over the valencies of all vertices, which must equal twice the number of edges.
We thus have
\bea
2\Delta +4l +2d = 2E = 2(2d+I), 
\eea
and hence we have the relation
\bea
\Delta = d+I-2l.\label{init2}
\eea

We can use \eref{init1} and \eref{init2} to find a bound on the number of faces
$d$ for Nakamura graphs of any genus $g$ and number of poles $n$. Using the equations to eliminate
$l$, we write
\bea
2(2g-2+n) - d = (\Delta + I).\label{init3}
\eea
The constants $\Delta$ and $I$ are always non-negative integers, so $d$ is bounded from above by $d_{max}$,
where 
\boxeq{
d_{max} := 2(2g-2+n) = 2|\chi|. \label{defdmax}
}
This is the maximum number of faces of a Nakamura graph of genus $g$ with $n$ fixed points.
To find Nakamura graphs computationally, it is helpful to first fix $|\chi|$ and then to find all the 
graphs of genus $g, n$ such that $|\chi|= (2g-2+n)$.

We can eliminate the number of internal edges $I$ from \eref{init1} and \eref{init2} to find a relation
between the branching number $\Delta$, the number of zeros $l$, and the Euler characteristic $|\chi|$:
\boxeq{
\Delta = |\chi| - l.\label{defDelta}
}
As $\Delta\geq0$, this equation gives us a bound on the number of zeroes of a Nakamura graph.
Since a Nakamura graph always has at least one zero, we have the bounds on the number of zeros of a Nakamura graph, 
\boxeq{
1 \leq l \leq |\chi|.\label{rangel}
}

The dimension of a cell associated to a Nakamura graph was derived in \cite{Nak}.
For a given graph with $l$ zeroes, $d$ faces and $n$ poles, 
we have $d$ width parameters.
The widths $b_k^{(i)}$ of the faces  bordering a given pole $P_i$ satisfy  a relation $\sum_k b_k^{(i)} = r_i$.
These residue relations specify $(n-1)$ independent constraints on the strip widths (since we have the total conservation equation $\sum_{i=1}^{n}r_i=0$).
There are $(l-1)$ real parameters corresponding to the independent time coordinates labelling the positions of the zeroes, modulo the overall time translation
symmetry.
So we can see that the dimension of the cell in moduli space corresponding to a Nakamura
graph is $(l-1)+d-(n-1)$. The above equations can be rearranged to show that the real dimension of
a cell is
\bea
\mbox{dim}_{\mathbb{R}}(\cC) = l+d-n= 6g-6+2n - (2\Delta+I).
\eea
This means that for a given genus and number of points $n$, the top dimension of the moduli space
of graphs is $6g-6+2n$, and the codimension of a given cell is 
\bea
\mbox{dim}_{\mathbb{R}}(\cM_{g,n}) - \mbox{dim}_{\mathbb{R}}(\cC) =2\Delta + I.
\eea

\section{Nakamura graphs as dessins d'enfants}\label{sec:belyinak}

In Section \ref{sec:review}, it was discussed that for a given $g$, $n$, and set of real numbers $r_1,\ldots, r_n$ that sum to zero, there is a cell decomposition of $\cM_{g,n}$, the moduli space of inequivalent Riemann surfaces, in which each cell is specified by a Nakamura graph $\cG$. Different points in the same cell in moduli space correspond to inequivalent Riemann surfaces with the same Nakamura graph but different Giddings-Wolpert differentials.

In this section we introduce a method to categorise the cells in moduli space by classifying the possible Nakamura graphs using
permutation groups and {\it dessins d'enfants}. 
We first review the notion of a dessin and discuss two distinct prescriptions for converting graphs into dessins. 
In each prescription, we show that there is a unique equivalence class of permutation triples 
corresponding to each Nakamura graph. We show that the necessary defining properties of Nakamura graphs can be 
encapsulated in the language of permutation groups, and hence equivalence classes of permutation triples 
can be used to catalogue the cells in moduli space.

\subsection{Review: dessins d'enfants}

A dessin d'enfant is a cyclically-ordered graph (a ribbon graph) that is also {\it bipartite}: each graph vertex
is coloured in black or white in such a way that black vertices only connect directly to white vertices, and white vertices
only connect to black vertices.
Given a bipartite graph with $r$ edges, we can assign an arbitrary labelling of $r$ objects to each edge,
such as the integers $\{1,2, \ldots r\}$.
Each vertex can be associated to a permutation cycle in $S_r$, representing the cyclic ordering of the edges connecting
to the vertex. 
As each edge connects to exactly one black and one white vertex, each integer in $\{1,2,\ldots, r\}$ appears in exactly
one cycle corresponding to a black vertex and in exactly one cycle corresponding to a white vertex. We can collate all the 
cycles corresponding to the black vertices to a single permutation $\sigma_1 \in S_r$, and likewise collate all the cycles corresponding to the white vertices to a permutation $\sigma_2\in S_r$.
The pair of permutations $(\sigma_1, \sigma_2)$ is enough to completely reconstruct the original dessin. 
In addition, we can introduce a third permutation $\sigma_3$, defined by the relation
\bea
\sigma_1 \sigma_2 \sigma_3 = 1.
\eea
This third permutation describes the structure of the faces of the dessin. 

A triple of $S_r$ permutations determines a unique dessin, but there will be other triples in $S_r$ that specify the same graph,
due to the arbitrariness of our original choice of labelling of the edges. 
This relabelling symmetry is described by an equivalence relation of conjugation on the permutation triples:
two triples $(\sigma_1, \sigma_2, \sigma_3)$ and $(\sigma_1', \sigma_2', \sigma_3')$ are equivalent if there exists some permutation $\gamma\in S_r$ acting on the edge labels of the graph such that
\bea
(\sigma_1', \sigma_2', \sigma_3') = (\gamma\sigma_1\gamma^{-1}, \gamma\sigma_2\gamma^{-1}, \gamma\sigma_3\gamma^{-1}).
\eea
This means that each dessin d'enfant with $r$ edges corresponds to an equivalence class of $S_r$ permutations under conjugation by $S_r$.

An automorphism of a dessin d'enfant is a mapping of the edges and vertices of the graph into itself 
such that the connections of the edges to the vertices, the colours of the vertices, and the cyclic ordering of the edges at the vertices are all preserved.
For a dessin described by a triple, these mappings are precisely the subgroup of $S_r$ consisting of elements $\gamma$ that satisfy
\bea
(\gamma^{-1}\sigma_1\gamma, \gamma^{-1}\sigma_2\gamma, \gamma^{-1}\sigma_3\gamma) = (\sigma_1, \sigma_2, \sigma_3).
\eea

A dessin d'enfant is said to be {\it clean} if each white vertex is bivalent (has valency two).
Any ribbon graph can be converted into a clean bipartite graph by colouring all the vertices in black and introducing
a new white vertex on each edge. The new graph has twice as many edges as the original graph.
This means that it is always possible to associate a dessin, and hence an equivalence class of permutation triples, to a ribbon graph. An example of a clean dessin d'enfant is included below on the right of Figure \ref{fig:naktodessin}.

\subsection{Nakamura graphs as \texorpdfstring{$S_{4d+2I}$}{S 4d+2I}-triples }

\begin{figure}[t]
\begin{center}
\includegraphics[width=0.7\textwidth]{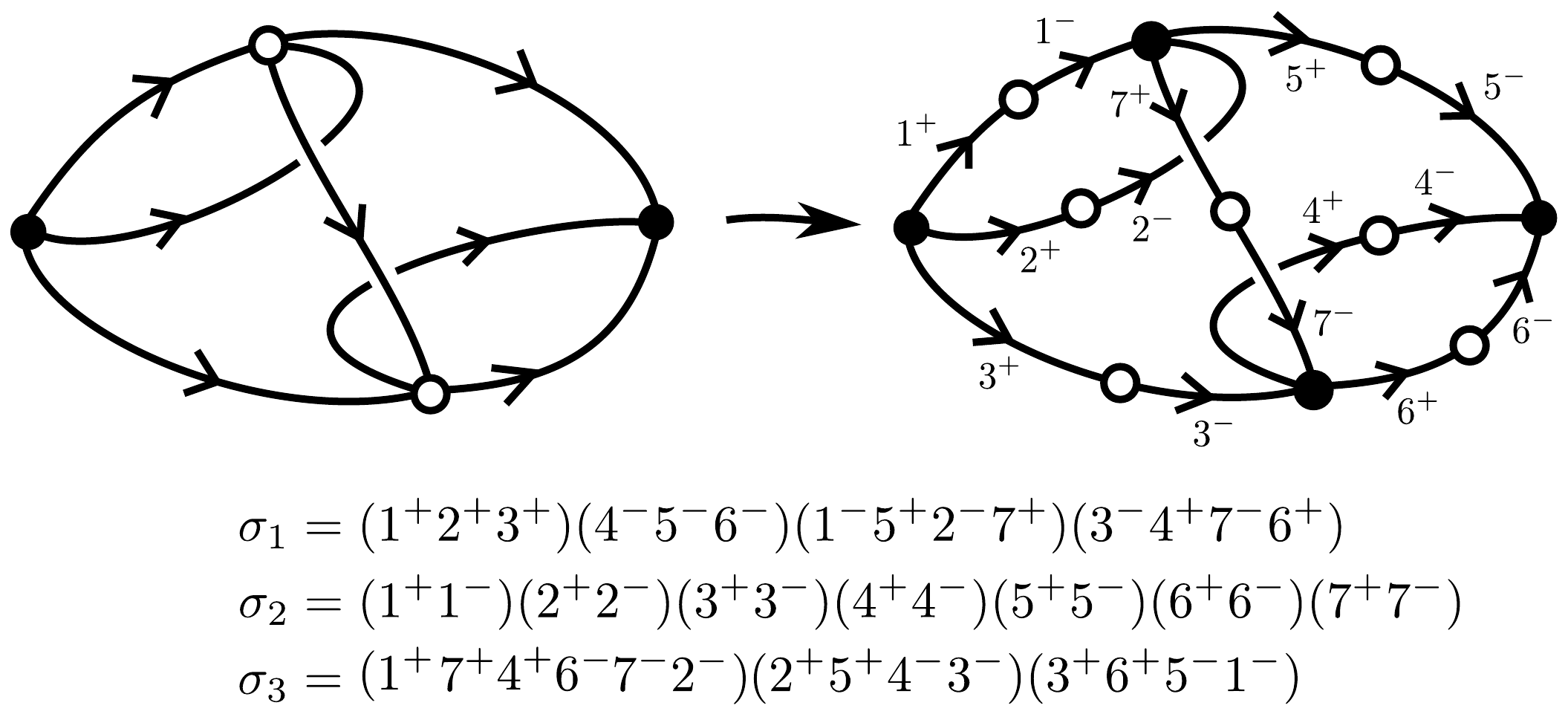}
\caption{Converting a Nakamura graph with $d=3$ and $I=1$ to a dessin d'enfant described by an $S_{4d+2I}$ triple.}
\label{fig:naktodessin}
\end{center}
\end{figure}
Nakamura graphs are oriented ribbon graphs satisfying a list of properties given in Section \ref{sec:props}.
Every graph has $d$ edges connecting to positive poles, $d$ edges connecting to negative poles, and $I$ edges
connecting only to zeroes, and so each graph has $2d+I$ edges in general.
As Nakamura graphs are not bipartite in general, they can only be described by permutation triples after cleaning
(introducing new vertices). Cleaning a graph doubles the number of edges of a graph, so a Nakamura graph 
dessin has $4d+2I$ edges in general. This means that every Nakamura graph can be described as a triple of 
$S_{4d+2I}$ permutations with overall conjugation equivalence by $S_{4d+2I}$. The poles and zeroes of the 
Nakamura graph all correspond to the black vertices.

We can fix some of the conjugation symmetry of Nakamura graphs by taking a canonical choice of the labelling
of the edges.
The number of edges of a dessin originating from cleaning a Nakamura graph is always even, so we can choose to label the
edges by $\{1^+, 1^-, 2^+, 2^-, \ldots, (2d+I)^+, (2d+I)^-\}$. Each edge of a Nakamura
graph has an orientation, and so each edge of the cleaned Nakamura graph has an orientation.
There are $d$ edges connected to the incoming poles, and $d$ edges connected to outgoing poles,
so we can label the edges connecting to incoming poles with the integers $\{1^+,\ldots, d^+\}$, and the edges going into the
outgoing poles by $\{(d+1)^-, \ldots, 2d^-\}$. 
These edges connect to bivalent white vertices: we can label the other connecting edges with the labels $\{1^-, \ldots, d^-\}$ and $\{(d+1)^+, \ldots, 2d^+\}$ such that each white vertex connects to edges labelled with the same integer but with different superscripts.
We label the edges connecting between the zeroes by integers from $(2d+1)^\pm$ to $(2d+I)^\pm$,
assigning integers with a $`+'$-superscript to the edges oriented from a black vertex to a white vertex, and a $`-'$-superscript to the edges 
oriented from white to black, such that each white vertex connects to edges labelled with the same integer but with different superscripts.

Each of the $l$ zeroes of a Nakamura graph connects to edges with cyclically alternating orientation. This is reflected in the structure of their
corresponding cycles; each cycle associated to a zero consists of a string of alternating $+$ and $-$-superscripted labels. These cycles
appear in the permutation $\sigma_1$. Also, all cycles in the permutation $\sigma_2$ are 2-cycles of the form $(i^+i^-)$ for some $i\in\{1,\ldots, 2d+I\}$.
The permutation $\sigma_3$ consists of $d$ cycles, corresponding to the $d$ faces of the ribbon graph. Each cycle in $\sigma_3$ consists of a string of consecutive 
$+$-superscripted integers, followed by a string of $-$-superscripted integers, which reflects the fact that each face is holomorphic to a strip.
An example of a dessin with this kind of labelling arising from a Nakamura graph is given above in Figure \ref{fig:naktodessin}.

With this choice of labelling, we can always uniquely decompose the permutation $\sigma_1$ into 
\bea \sigma_1=\sigma_+\sigma_-\sigma_Z,\eea 
where $\sigma_+$ describes the incoming poles and acts on the set $\{1^+,2^+,\ldots, d^+\}$, 
$\sigma_-$ describes the outgoing poles and acts on the set $\{(d+1)^-,(d+2)^-,\ldots, 2d^-\}$, 
and $\sigma_Z$ describes the zeroes of the graph and acts on the remaining $2d+2I$ edges. 
The permutation $\sigma_2$ can be written
\bea
\sigma_2 &=& \prod_{i=1}^{2d+I}(i^+i^-),
\eea
and, schematically, $\sigma_3$ is of the form
\bea
\sigma_3 &=& \prod_{k=1}^d \alpha_k,  \qquad \alpha_k=(i_1^+,i_2^+,\ldots,i_p^+,j_1^-,j_2^-,\ldots,j_q^-).
\eea
Our choice of labelling `breaks' the $S_{4d+2I}$ conjugation symmetry down to a smaller subgroup.
Two permutation descriptions of a graph $(\sigma_+, \sigma_-, \sigma_Z, \sigma_2)$ and $(\sigma_+', \sigma_-', \sigma_Z', \sigma_2')$ with the above conventions for labellings are equivalent if there is some $\gamma\in S_{4d+2I}$ satisfying
\bea
(\gamma^{-1}\sigma_+\gamma,\ \gamma^{-1}\sigma_-\gamma,\ \gamma^{-1}\sigma_Z\gamma,\  \gamma^{-1}\sigma_2\gamma, ) = (\sigma'_+, \sigma'_-, \sigma'_Z, \sigma'_2 ).
\eea
If we wish to find which conventionally-labelled permutation triples are equivalent, we need only consider 
equivalence of the triples under those permutations in an $S_{4d+2I}$ subgroup that
preserve the required forms of $\sigma_+$, $\sigma_-$, $\sigma_Z$, and $\sigma_2$ separately. 
We can thus just consider conjugation of conventionally-labelled permutation tuples under 
\bea
\gamma\in (S_{d}\times S_d\times S_{2d+2I}) \cap S_{2d+I}[S_2],
\eea 
where $S_{2d+I}[S_2]$ is the wreath product.

\begin{figure}[t]
\begin{center}
\includegraphics[width=0.40\textwidth]{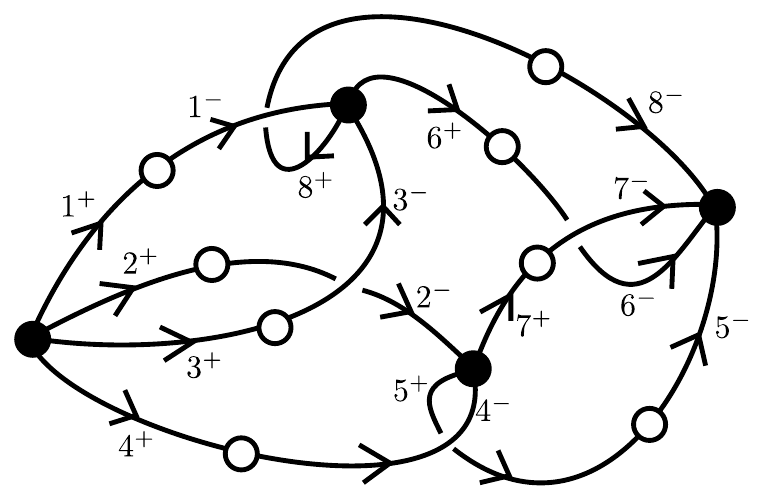}
\caption{The $S_{4d+2I}$ dessin associated to a Nakamura graph with an automorphism group of order 4.}
\label{fig:naiveeg}
\end{center}
\end{figure}

The automorphisms of a Nakamura graph are the ribbon graph automorphisms which also preserve the orientation of the edges.
In particular, this means that Nakamura graph automorphisms map positive poles to positive poles, negative poles to negative 
poles, and zeroes to zeroes. Automorphisms are allowed to permute poles of the same sign.
In the $S_{4d+2I}$ picture, we can decompose the permutation $\sigma_1=\sigma_+\sigma_-\sigma_Z$. For the orientations of 
the graph to be preserved, the automorphisms 
must preserve these three constituent permutations separately. Hence the automorphism group of a Nakamura graph in the
$S_{4d+2I}$ picture is a subgroup  Aut$(\{\sigma_i\}) \subset S_{4d+2I}$ such that
$\gamma\in \text{Aut}(\{\sigma_i\})$ if 
\bea
(\gamma^{-1}\sigma_+\gamma,\ \gamma^{-1}\sigma_-\gamma,\ \gamma^{-1}\sigma_Z\gamma,\  \gamma^{-1}\sigma_2\gamma, ) = (\sigma_+, \sigma_-, \sigma_Z, \sigma_2 ).
\eea
(The condition $\gamma^{-1}\sigma_3\gamma=1$ is automatically satisfied by the fact that $\sigma_1\sigma_2\sigma_3=1$.)

An example of a conventionally-labelled dessin d'enfant in the $S_{4d+2I}$ description is given in Figure \ref{fig:naiveeg}.
This graph is described by a triple of permutations acting on the set of 16 elements $\{1^+, 1^-, \ldots, 8^+, 8^-\}$:
\bea
\sigma_1 = (1^+2^+3^+4^+)(5^-6^-7^-8^-)(1^-6^+3^-8^+)(2^-7^+4^-5^+), \nonumber \\
\sigma_2 = (1^+1^-)(2^+2^-)(3^+3^-)(4^+4^-)(5^+5^-)(6^+6^-)(7^+7^-)(8^+8^-), \nonumber \\
\sigma_3 = (1^+8^+7^-2^-)(2^+5^+8^-3^-)(3^+6^+6^-4^-)(4^+7^+6^-1^-). \label{eq:S4d2Itriple}
\eea
The black vertices correspond to $\sigma_1$, the white vertices correspond to $\sigma_2$, and the 
faces of the graph correspond to $\sigma_3$.
The automorphism group of this graph is isomorphic to $\mathbb{Z}_4$, and is generated by 
\bea
\gamma = (1^+2^+3^+4^+)(5^+6^+7^+8^+)(1^-2^-3^-4^-)(5^-6^-7^-8^-).
\eea

\subsection{Nakamura graphs as \texorpdfstring{$S_{2d+2I}$}{S 2d+2I}-triples}

\begin{figure}[t]
\begin{center}
\includegraphics[width=0.7\textwidth]{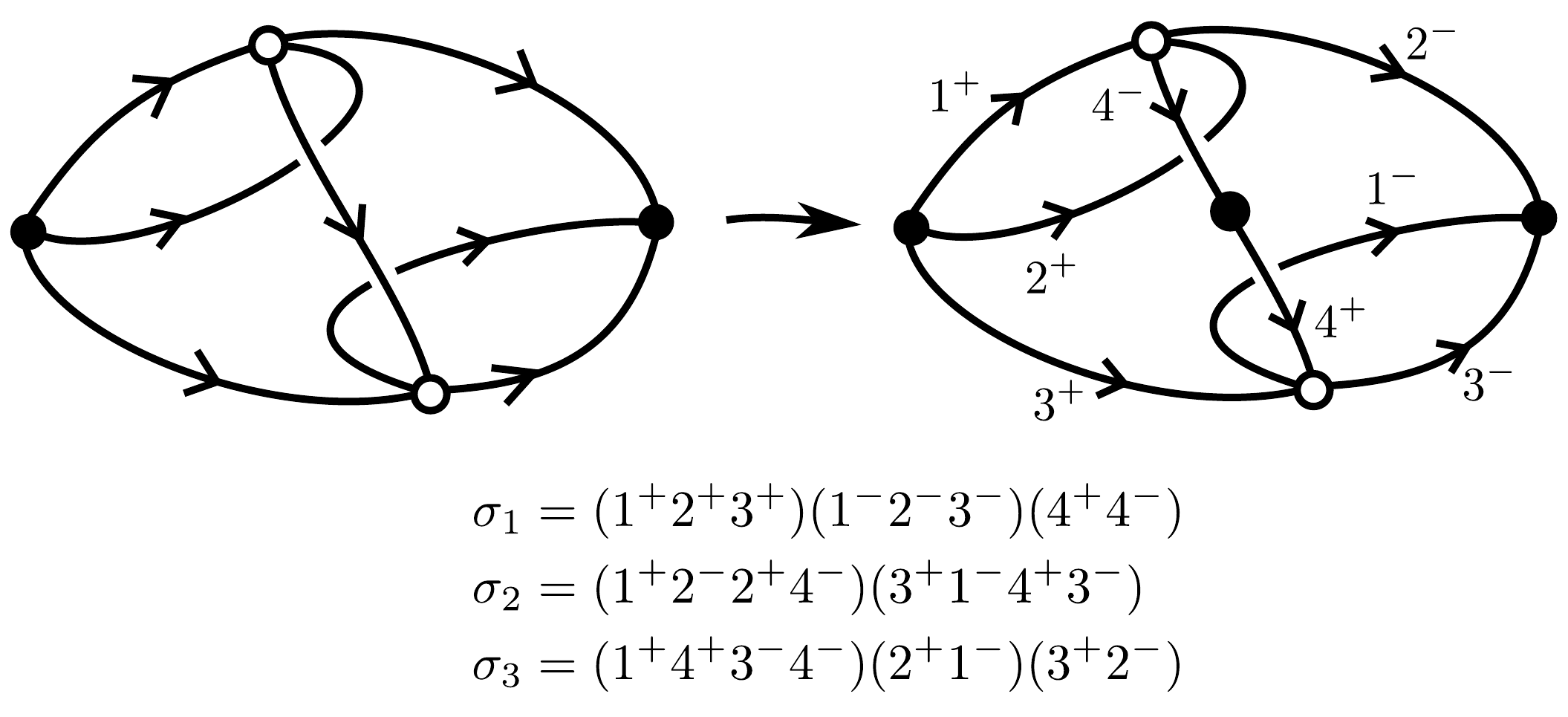}
\caption{Converting a Nakamura graph with $d=3$ and $I=1$ to a dessin d'enfant described by an $S_{2d+2I}$ triple.}
\label{fig:naktoecon}
\end{center}
\end{figure}
The description of a general Nakamura graph in terms of a triple of permutations is possible because the graph 
can be made into a clean bipartite graph by adding extra vertices. Without the addition of extra vertices, 
Nakamura graphs are not bipartite in general. However, the property that no pole connects to another pole allows us to find a permutation tuple description requiring fewer labelled edges, and hence requiring permutation groups of smaller degree.

Starting from a Nakamura graph, colour the poles in black and the zeroes in white. Subdivide only the internal edges connecting zeroes to zeros
by adding in extra vertices. As there are no edges connecting poles to poles, this graph must be bipartite. 
Label the edges going out of the incoming poles by $\{1^+,\ldots, d^+\}$, and the edges going into the outgoing poles by $\{1^-, \ldots, d^-\}$.
Label the edges bordering each zero with integers from $(d+1)^\pm$ to $(d+I)^\pm$, such that the edges oriented towards a zero are assigned a 
$-$-superscripted integer, and the edges oriented away from a zero are assigned the corresponding $+$-superscripted integer.

As in the $S_{4d+2I}$ description, this bipartite graph can be described by a triple of permutations $\sigma_1$, $\sigma_2$, and $\sigma_3$ satisfying $\sigma_1\sigma_2\sigma_3=1$. The permutation $\sigma_1$ describes the structure of the graph at the poles and at the new vertices
added in the internal edges, $\sigma_2$ describes the graph at the zeroes, and $\sigma_3$ describes the faces of the graph.
We can decompose $\sigma_1$ into three permutations with $\sigma_1=\sigma_+\sigma_-\sigma_I$, where $\sigma_+$ acts on
$\{1^+, \ldots, d^+\}$ and describes the 
incoming poles, $\sigma_-$ acts on  $\{1^-, \ldots, d^-\}$ and describes the outgoing poles, and $\sigma_I=\prod_{i=d+1}^{d+I}(i^+i^-)$
describes the $I$ internal edges. The permutation $\sigma_2$ now describes the $l$ zeroes, and so each of the $l$ cycles consists of a 
string of alternating $+$, $-$-superscripted labels. As in the $S_{4d+2I}$ description, $\sigma_3$ is of the form
\bea
\sigma_3 &=& \prod_{k=1}^d \alpha_k,  \qquad \alpha_k=(i_1^+,i_2^+,\ldots,i_p^+,j_1^-,j_2^-,\ldots,j_q^-).
\eea
This new descriptions requires only $2d+2I$ labelled edges for each graph.
The choice of labelling of the edges allows us to state that two tuples of permutations $(\sigma_+, \sigma_-, \sigma_I, \sigma_2)$ 
and $(\sigma_+', \sigma_-', \sigma_I', \sigma_2')$ are equivalent if they are conjugate by a permutation $\gamma$, where
\bea
\gamma \in S_d \times S_d \times S_I[S_2].
\eea
An example of a dessin described by an $S_{2d+2I}$ triple is given in Figure \ref{fig:naktoecon}.

The automorphisms of a Nakamura graph in the $S_{2d+2I}$ picture are the automorphisms of the $S_{2d+2I}$ dessin that 
preserve the orientation of the edges in the dessin. The permutation $\sigma_1$ decomposes as $\sigma_+\sigma_-\sigma_I$, and so the automorphisms of the graph in this picture are the subgroup $\text{Aut}(\{\sigma_i\}) \subset S_{2d+2I}$ such that $\gamma\in \text{Aut}(\{\sigma_i\})$ if 
\bea
(\gamma^{-1}\sigma_+\gamma,\ \gamma^{-1}\sigma_-\gamma,\ \gamma^{-1}\sigma_I\gamma,\  \gamma^{-1}\sigma_2\gamma, ) = (\sigma_+, \sigma_-, \sigma_I, \sigma_2 ).
\eea

\begin{figure}[t]
\begin{center}
\includegraphics[width=0.40\textwidth]{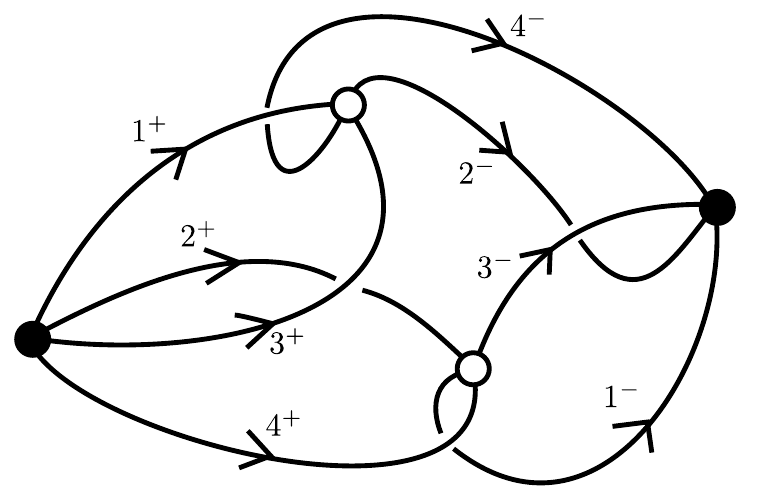}
\caption{The $S_{2d+2I}$ dessin associated to a Nakamura graph with an automorphism group of order 4.}
\label{fig:econeg}
\end{center}
\end{figure}

The example of a Nakamura graph with automorphism group of order four given in the previous section
can be described in the $S_{2d+2I}$ picture.
The graph drawn in Figure \ref{fig:econeg} is described by a triple of permutations acting on the set of
8 elements $\{1^+,1^-,\ldots, 4^+, 4^-\}$:
\bea
\sigma_1 &=& (1^+2^+3^+4^+)(1^-2^-3^-4^-) \nonumber \\
\sigma_2 &=& (1^+2^-3^+4^-)(2^+3^-4^+1^-) \nonumber \\
\sigma_3 &=& (1^+3^-)(2^+4^-)(3^+1^-)(4^+2^-) \label{eq:S2d2Itriple}
\eea
The automorphism group of the $S_{2d+2I}$ dessin is necessarily isomorphic to the automorphism group of
the $S_{4d+2I}$ dessin, as they are both descriptions of the same Nakamura graph.
In this case, the automorphism group $\mathbb{Z}_4$ is generated by 
\bea
\gamma = (1^+2^+3^+4^+)(1^-2^-3^-4^-).
\eea

\subsection{From permutation triples to cells in moduli space}\label{sec:triplestocells} 

Given a Nakamura graph with $d$ faces and $I$ internal edges, it is always possible to construct a triple of permutations from the group $S_{4d+2I}$ or $S_{2d+2I}$ that describes the graph.
Not every triple of permutations in these groups corresponds to a Nakamura graph, though.
For a given triple of permutations to describe a Nakamura graph, it must satisfy a particular set of conditions.

A triple of $S_{4d+2I}$ permutations $(\sigma_1, \sigma_2, \sigma_3)$ specifies a conventionally-labelled Nakamura graph
if it satisfies the following properties:
\begin{itemize}
\item
The subgroup generated from $\sigma_1$ and $\sigma_2$ acts transitively on $X=X^+\cup X^-$, where 
\bea  
X^+= \{ 1^+, 2^+, \ldots, (2d+I)^+ \}, \\
X^- = \{1^-, 2^-, \ldots, (2d+I)^-\}.
\eea
(This is the condition that a Nakamura graph is connected.)
\item 
The permutation $\sigma_1$ can be written as 
\bea
\sigma_1 = \sigma_+ \s_- \s_Z,
\eea
where $\s_+$, $\s_-$ and $\s_Z$ are disjoint, and:
	\begin{itemize}
	\item $\s_+$ acts on $\{1^+, 2^+, \ldots, d^+\}$ and fixes all other elements,
	\item $\s_-$ acts on $\{(d+1)^-, (d+2)^-, \ldots, 2d^-\}$ and fixes all other elements,
	\item $\s_Z$ has no cycle of length less than 4, $\s_Z(X^+)=X^-$, and $\s_Z(X^-)=X^+$.
	\end{itemize}
(This is the condition that a Nakamura graph decomposes into positive poles, negative poles, and zeroes, and that the orientations
of the connecting edges are outgoing, incoming and alternating respectively.)
\item $\s_2 = (1^+1^-)(2^+2^-)\ldots ((2d+I)^+(2d+I)^-).$
(This is the condition that the $S_{4d+2I}$ dessin is clean.)
\item
The permutation $\s_3$ decomposes into $d$ disjoint cycles as $\sigma_3 = \sigma_3^{(1)} \sigma_3^{(2)}\ldots \sigma_3^{(d)}$, where for each $\sigma_3^{(i)}$
\bea
|\sigma_3^{(i)}(X^+)\cap X^-| = 1 = |\sigma_3^{(i)}(X^-)\cap X^+|.
\eea
(This is the condition that each disjoint cycle in $\sigma_3$ corresponding to a face of the graph is of the form $(++\ldots+--\ldots-)$, and so corresponds to a strip.)
\item
For any sequence of non-negative integers $(k_1, k_2, \ldots, k_r)$ and some $i^+ \in X^+$, if all the elements of the sequence
\bea
i^+, \quad
\sigma_2\sigma_Z^{2k_1+1}(i^+), \quad
\sigma_2\sigma_Z^{2k_1+1}\sigma_2\sigma_Z^{2k_2+1}(i^+), \quad  \ldots
\eea
are contained in $X^+$, then this sequence has no repeated element.
(This condition forbids closed oriented loops on the graph, and permits time orderings to be assigned to the zeroes
of the graph.)
\end{itemize}
\vspace{5mm}

Similarly, a triple of $S_{2d+2I}$ permutations $(\sigma_1, \sigma_2, \sigma_3)$ specifies a conventionally-labelled Nakamura graph if it satisfies the following properties:
\begin{itemize}
\item 
The edges can be assigned labels from the set $X=X^+\cup X^-$, where 
\bea  
X^+= \{ 1^+, 2^+, \ldots, (d+I)^+ \}, \\
X^- = \{1^-, 2^-, \ldots, (d+I)^-\}.
\eea
\item 
The subgroup generated from $\sigma_1$ and $\sigma_2$ acts transitively on $X$.
\item 
The permutation $\sigma_1$ can be written as 
\bea
\sigma_1 = \sigma_+ \s_- \s_I,
\eea
where $\s_+$, $\s_-$ and $\s_I$ are disjoint, and:
	\begin{itemize}
	\item $\s_+$ acts on $\{1^+, 2^+, \ldots, d^+\}$ and fixes all other elements,
	\item $\s_-$ acts on $\{1^-, 2^-, \ldots, d^-\}$ and fixes all other elements,
	\item $\s_I = ((d+1)^+(d+1)^-)\ldots((d+I)^+(d+I)^-)$.
	\end{itemize}
\item 
The permutation $\s_2$ has no cycle of length less than 4, $\s_2(X^+)=X^-$, and $\s_2(X^-)=X^+$.
\item 
The permutation $\s_3$ decomposes into $d$ disjoint cycles as $\sigma_3 = \sigma_3^{(1)} \sigma_3^{(2)}\ldots \sigma_3^{(d)}$, where for each $\sigma_3^{(i)}$
\bea
|\sigma_3^{(i)}(X^+)\cap X^-| = 1 = |\sigma_3^{(i)}(X^-)\cap X^+|.
\eea
\item
For any sequence of non-negative integers $(k_1, k_2, \ldots, k_r)$ and some $i^+ \in \cal{I}^+$, where
${\cal I}^+ = \{ (d+1)^+,\ldots, (d+I)^+\}$, if all the elements of the sequence
\bea
i^+, \quad
\sigma_I\sigma_Z^{2k_1+1}(i^+), \quad
\sigma_I\sigma_Z^{2k_1+1}\sigma_I\sigma_Z^{2k_2+1}(i^+), \quad  \ldots
\eea
are contained in $\cal{I}^+$, then this sequence must not have a repeated element.
\end{itemize}

\section{Nakamura graphs as Hurwitz classes}\label{sec:hurwitz}

In the previous section we introduced two methods of describing Nakamura graphs with 
triples of permutations which multiply to the identity by converting the Nakamura graphs to bipartite graphs
with extra vertices. These triples of permutations are elements of either $S_{2d+2I}$ or $S_{4d+2I}$, 
where $d$ is the number of strips (faces) of a graph and $I$ is the number of internal edges in the graph
connecting zeroes to zeroes. However, in this description, the conditions that a general permutation
triple must satisfy to be a Nakamura graph are rather cumbersome, and can be tricky to check computationally.

In this section we present a new description of a Nakamura graph in terms of a tuple of $m+2$ permutations in $S_d$
which multiply to the identity, where $m\leq l$, and $l$ is the number of zeroes of the graph.
This approach has two main advantages over the triples description:
the necessary permutation group $S_d$ is smaller than $S_{2d+2I}$ or $S_{4d+2I}$, and
the set of conditions that a generic tuple must satisfy to give a Nakamura graph is much simpler.
Both conditions mean that it is easier to implement Nakamura graphs computationally with the group 
$S_d$ than with the groups $S_{2d+2I}$ or $S_{4d+2I}$.

We begin this section with a review of {\it Hurwitz theory}, which describes how
equivalence classes of branched covers of Riemann surfaces correspond to equivalence classes 
of permutation tuples multiplying to the identity. We will call such an equivalence class of permutations
a {\bf Hurwitz class}.  More on this standard subject of algebraic topology 
can be found, for example, in \cite{ezell,hatcher,fulton} or in a physics context in \cite{grosstaylor,cmr}.  
 The equivalence classes of permutations triples discussed in Section \ref{sec:belyinak}
are examples of Hurwitz classes. We then discuss how to construct branched covers from a 
Riemann surface with a Giddings-Wolpert differential to an infinite cylinder, with the ramification points 
of the surface being exactly the poles and zeroes of the GW differential.
The Hurwitz class corresponding to this cover is an equivalence class of a tuple of $m+2$ permutations in $S_d$,
and contains enough information to reconstruct the Nakamura graph associated to the domain Riemann surface.

Each Hurwitz class corresponds to a single Nakamura graph, but a Nakamura graph may correspond to many
distinct Hurwitz classes. This makes it difficult to find the automorphism group of a Nakamura graph from
a generic Hurwitz class associated to the graph. To solve this issue, we introduce a new equivalence
relation on the set of Hurwitz classes - which we call {\bf slide-equivalence} - such that the equivalence classes
of this relation are in one-to-one correspondence with the Nakamura graphs. Within the slide-equivalence class
of any Nakamura graph, there is a unique canonical choice of a Hurwitz class - whose elements we call {\bf reduced tuples} -
that yields in a simple way the automorphism group of the associated graph.
This description gives a computationally powerful method of finding the Nakamura graphs and their automorphism groups.

\subsection{Review: Branched covers, Hurwitz classes, and Belyi maps}

\begin{figure}[t]
\begin{center}
\includegraphics[width=0.85\textwidth]{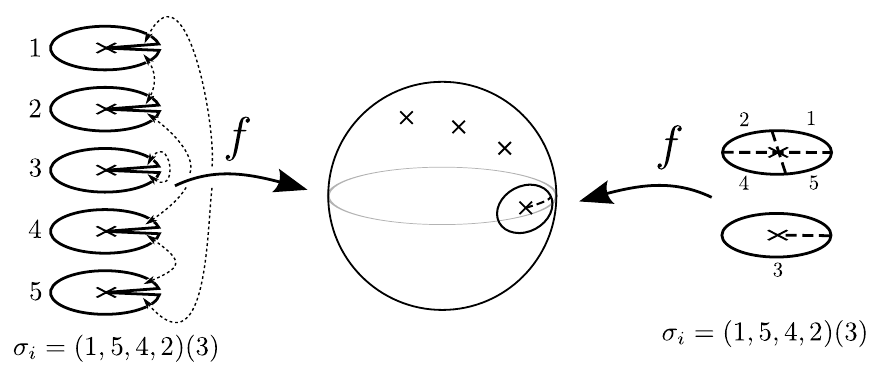}
\caption{The preimages of a cut disc on $S^2$ are a set of cut discs, whose gluing is specified by a permutation $\sigma_i$.}
\label{fig:branching2}
\end{center}
\end{figure}
A continuous surjective map $f: \Sigma \to S^2$ is a {\bf branched cover} of the Riemann sphere  if every point $Q$ on $S^2$ has some open neighbourhood $U_Q$ such that $f^{-1}(U_Q)$ is a collection of disjoint open sets, and on each set $f$ is topologically equivalent to the complex map $z \mapsto z^r$ for some positive integer $r$.
For most points on the sphere, there are $d$ preimages on the surface $\Sigma$, where $d$ is the degree of the map.
There is a finite set of points on the target space $S^2$ which each have fewer than $d$ preimages. These are
the {\bf branch points} of the map $f$.
Consider a point $Q$ on the surface $S^2$. 
If $Q$ is not a branch point, then for each of its preimages $P$ on $\Sigma$, there exist complex coordinate 
patches $z$ about $P$ and $w$ about $Q$ such that $f$ maps $z\mapsto w=z$. However,
if $Q$ is a branch point, then for at least one of its preimages $P$ there exist coordinate patches $z$ 
about $P$ and $w$ about $Q$ where $f$ maps $z \mapsto w= z^r$ for $r\geq2$. Such a point $P$ is 
called a {\bf ramification point} of the map $f$.
For a given branch point $Q$, each preimage $P_i$ of the branch point has an associated unique positive integer $r_i$ 
such that $f$ maps $z \mapsto w=z^{r_i}$ about that point. The tuple of integers $(r_1, r_2, \ldots )$ is
the {\bf ramification profile} of the branch point $Q$.

The neighbourhoods of ramification points can be described in terms of a gluing construction.
Take a disc around a branch point $Q$ with coordinates $|w|<1$, and cut the disc along the real interval
$w\in[0,1)$. The preimages of the cut disc on the surface $\Sigma$ are $d$ identical copies of the cut disc.
The cuts along the intervals can be identified to recover the neighbourhoods on $\Sigma$ around the ramification points.
If we choose a labelling of the cut discs with the integers $\{1,2,\ldots, d\}$, then the 
gluing of the cut discs corresponds to a mapping from the set $\{1,2,\ldots, d\}$ to itself: the lower edge of the cut on disc $i$ is glued to the upper edge of the cut on disc $\sigma(i)$. This gluing is shown on the left of Figure \ref{fig:branching2}.
Each cut disc is biholomorphic to a `wedge' of a disc subtending an angle $2\pi/r$ for some $r$, as can be seen on
the right of Figure  \ref{fig:branching2}.

\begin{figure}[t]
\begin{center}
\includegraphics[width=0.7\textwidth]{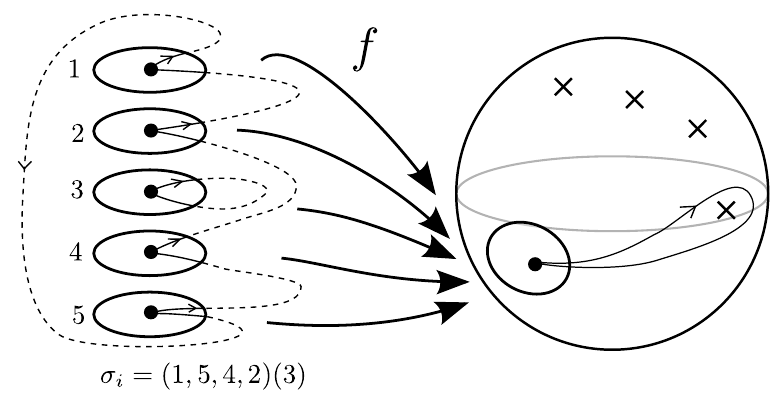}
\caption{The target space $S^2$ is drawn on the right and the $d$ preimages on the surface $\Sigma$ of a disc about a marked unbranched point on the sphere are drawn on the left.
The preimages of a loop drawn around one of the branch points on the sphere are a set of 
trajectories connecting the $d$ labelled preimages of the marked point on $\Sigma$, and this specifies a permutation in $S_d$.}
\label{fig:branching}
\end{center}
\end{figure}

There is another way of arriving at the permutation description of branch points by considering the preimages
of loops on the target space $S^2$.
Choose a marked unbranched point on the sphere, and label its preimages with integers from $1$ to $d$.
For each of the $l$ branch points on the sphere, draw a directed closed path starting and ending on the marked point,
which can be contracted to a neighbourhood of the branch point without passing through a branch point.
The preimages of each of the $l$ directed loops on the sphere are directed closed paths on the Riemann surface $\Sigma$
which connect the $d$ distinct labelled preimages of the marked point. Each branch point gives a bijective mapping 
from the set $\{1,\ldots d\}$ to itself which we obtain by following the paths of the preimages of the loops. 
We associate a permutation $\sigma_i \in S_{d}$, $i=1,\ldots m$ to each branch point of the map $f$. 
On the sphere, the path constructed by following all $m$ loops around is contractible.
Hence, the permutations $\sigma_1, \ldots, \sigma_m$ multiply together to give the identity,
\bea
\sigma_1\sigma_2\ldots \sigma_m = 1.
\eea
The permutation tuple $(\sigma_1, \sigma_2, \ldots \sigma_m)$ describes the branching profile of a branched
cover $f$ from a Riemann surface $\Sigma$ on to the sphere $S^2$. This is demonstrated in Figure \ref{fig:branching}.

There is an arbitrariness in the way we label the preimages of the marked point from $1$ to $d$:
any relabelling of these points yields the same branching profile. Hence, we consider two permutation tuples
to be equivalent if there is a permutation $\gamma \in S_d$ which conjugates one sequence to the other. That is,
the tuples $(\sigma_1, \ldots, \sigma_m)$ and $(\sigma'_1, \ldots, \sigma'_m)$ are equivalent if
\bea
(\sigma'_1, \ldots, \sigma'_m) = (\gamma \sigma_1 \gamma^{-1}, \ldots, \gamma \sigma_m\gamma^{-1} ).
\eea
We call an equivalence class of tuples under conjugation a {\bf Hurwitz class}.

There is also a notion of equivalence of branched coverings in terms of bijective maps. 
Two branched covers of the sphere $f$ and $f'$ are equivalent if there exists
some homeomorphism $\phi : \Sigma \rightarrow \Sigma$ such that $f' = f \circ \phi$. 
In other words, $f$ and $f'$ are equivalent if the following diagram commutes:
\bea 
&&  ~~~ \Sigma  ~~~~  \xlongrightarrow { \phi }  ~~~~~~  \Sigma \cr 
&&   ~~~ f' \searrow  ~~~~~~~ \swarrow f  \cr 
&&  ~~~~~~~~~~~ ~~ S^2
\eea
This definition of equivalence coincides with the $S_d$ conjugation equivalence: {\it two branched covers
of a Riemann surface are equivalent if they have the same Hurwitz class.}
\begin{figure}[t]
\begin{center}
\includegraphics[width=0.85\textwidth]{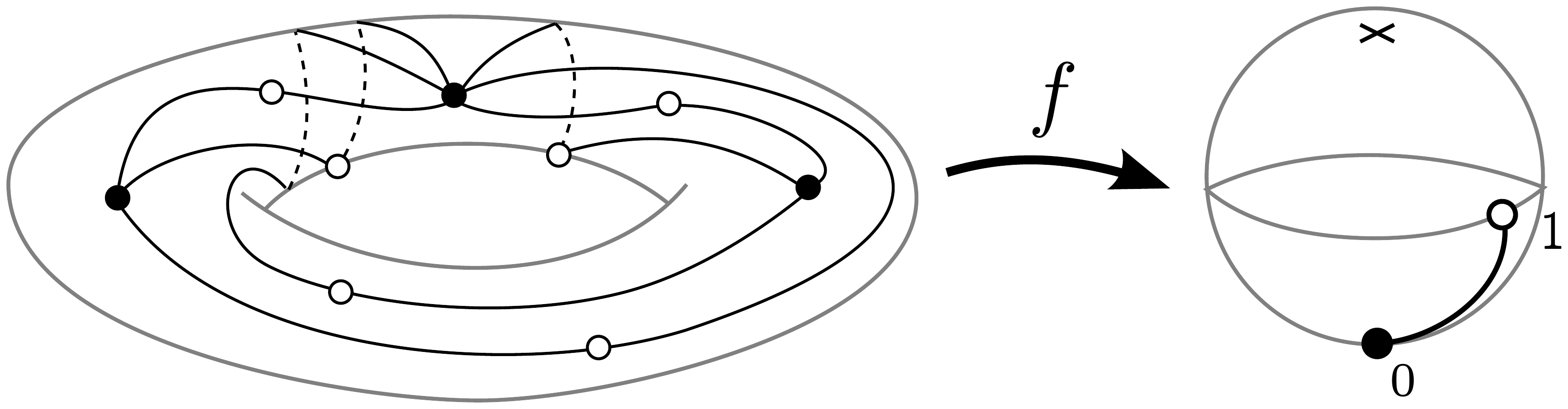}
\caption{Any dessin d'enfant on a Riemann surface can be realised as the preimage of $[0,1]$ on some branched covering of 
the sphere.}
\label{fig:p1}
\end{center}
\end{figure}
The genus of the covering surface can be expressed, according to the Riemann-Hurwitz relation,  in terms of the { \it branching numbers} $B_i = d - C_{ \sigma_i } $ of the branch points as 
\bea\label{RHrel} 
2g -2 = -2 d + \sum_{ i =1}^{ m } B_i 
\eea

Dessins d'enfants can be realised as branched coverings of the sphere. If we take a branched cover of the sphere 
with branch points located at $\{0,1,\infty\}$, and consider the real interval $[0,1]$ on the target sphere, then the preimage
of this interval on the Riemann surface is an embedded ribbon graph. Colouring the preimages of the point $w=0$ on the sphere 
in black and the preimages of $w=1$ in white, it can be seen that the embedded ribbon is bipartite and is therefore a dessin.
An example of a branched covering of the sphere generating a dessin d'enfant is shown in Figure \ref{fig:p1}.
If we choose a labelling of the $d$ preimages of the real interval, then we can find a Hurwitz class associated to the branched covering. This Hurwitz class coincides exactly with the defining equivalence class of a dessin d'enfant.
A branched covering of the sphere with three branch points is called a {\bf Belyi map}, and we call a representative element
of its associated Hurwitz class a {\bf Belyi triple}.
The Nakamura graph descriptions from Section \ref{sec:belyinak} are examples of Belyi triples which correspond to Belyi 
maps of degree $2d+2I$ or $4d+2I$.

\subsection{Nakamura graphs and branched coverings}

Consider a Riemann surface with a Giddings-Wolpert differential and embedded Nakamura graph.
The Nakamura graph partitions the surface into $d$ faces, each of which is holomorphic to an infinite complex strip, 
such as in Figure \ref{fig:pants}.
The zeroes of the differential lie on the boundaries of the strips, and the poles are located at the 
negative and positive infinities of the strips.
The surface can be reconstructed from the strips by a gluing of the edges determined by the Nakamura graph.

First, let us consider a Riemann surface with a GW differential in which the $d$ strips are of equal width $2\pi$. 
The strips can then be viewed as copies of a single template strip of width $2\pi$.
There is a trivial map from each of the $d$ worldsheet strips on to the target strip, in which all the preimages of a point on the 
target strip have the same time coordinate. On identifying the upper and lower edges of the target space strip, the map
extends to a branched covering from the surface onto the cylinder. 
All the real trajectories of the Nakamura graph are mapped on to a single infinite line on the cylinder, and all the zeroes 
are mapped on to this line. The positive (incoming) poles of the graph are mapped on to negative infinity, and the negative
(outgoing) poles of the graph are mapped on to positive infinity. The map has $m+2$ branch points, where $m\leq l$ is the number of distinct time coordinates of the zeroes. If the time coordinates of all the zeroes are distinct, then $m=l$.

An infinite cylinder of circumference $2\pi$ can be mapped bijectively to the Riemann sphere with the exponential map
$z \mapsto \exp z$. This means that the composition of the cylinder covering and the exponential map is a holomorphic 
branched covering $f$ of the Riemann sphere with $m+2$ branch points. 
The positive poles of the Nakamura graph map on to 0, the negative poles of the graph map on to $\infty$, and the remaining
$l$ zeroes map on to $m$ branch points along the real axis on the sphere.
The Giddings-Wolpert differential on the worldsheet is $\frac{df}{f}$.

Now consider a more general GW differential where the strips are no longer of equal width.
We can construct a bijective mapping from each strip onto a single template strip of width $2\pi$ in such a way that the preimages of a point on the template strip have the same time coordinate. However, this mapping will not be holomorphic in general.
Applying the exponential map to this template strip, we have a map $f$ from a general Riemann surface onto the sphere.
The GW differential cannot be written in the form $\frac{df}{f}$ in this more general case, but the map $f$ is still
a branched cover of the sphere, with ramification points at the poles and zeros of the differential.

\begin{figure}[t]
\begin{center}
\includegraphics[width=0.95\textwidth]{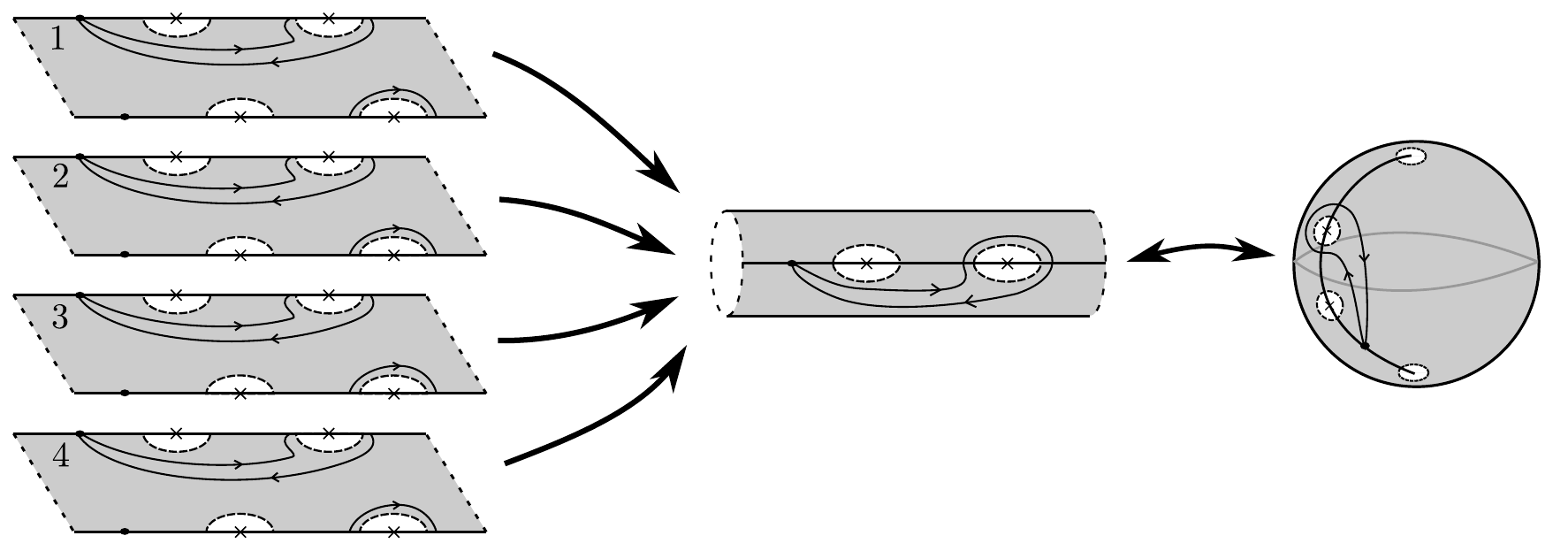}
\caption{Nakamura graph strips naturally form a branched cover of the cylinder and the sphere.}
\label{fig:cylindercoverloop}
\end{center}
\end{figure}

This branched cover of the sphere has an associated permutation tuple describing the branching.
We mark an unbranched point on the sphere and label the preimages of this point with the integers from $1$ to $d$. 
The preimage of a small loop starting and ending on this marked point that encloses a branch point on the Riemann sphere is a collection of closed paths connecting the labelled preimages of the unbranched point.
Each branch point determines a permutation $\sigma\in S_d$, and so the branched covering determines a tuple consisting
of $m+2$ permutations
\bea
(\sigma_+, \sigma_1, \sigma_2, \ldots, \sigma_m, \sigma_-),
\eea
that describes the gluing of the different strips. Here, the permutation $\sigma_+$ describes the branching about 0,
$\sigma_-$ describes the branching around $\infty$, and $\sigma_i$ describes the branching around the $i$th branch 
point on the real line. As this is a branched covering of the sphere, this set of permutations multiplies to one,
\bea
\sigma_+\sigma_1\sigma_2\ldots\sigma_m\sigma_- = 1.
\eea
There is also an overall conjugacy equivalence of the tuple due to the arbitrary choice of labelling of the $d$ inverse images
of the marked point, 
\bea
(\gamma \sigma_+ \gamma^{-1}, \gamma\sigma_1 \gamma^{-1}, \gamma\sigma_2\gamma^{-1}, \ldots, \gamma\sigma_m\gamma^{-1}, \gamma\sigma_-\gamma^{-1}) \sim (\sigma_+, \sigma_1, \sigma_2, \ldots, \sigma_m, \sigma_-),
\eea
where $\gamma \in S_d$. This construction is shown in Figure \ref{fig:cylindercoverloop}, where the marked point is chosen to lie on
the real axis of the Riemann sphere, and the preimages of this point lie on the boundaries of the strips. For the case $m=l$, the Riemann-Hurwitz relation \eref{RHrel} can be written 
as 
\bea 
(2g-2) = -n + l + \Delta 
\eea 
This also follows from the previous discussion of Nakamura graph parameters in Section 
\ref{Nakpars}, in particular by eliminating $I$ from equations  \eref{init1} and \eref{init2}. 

The boundaries of the strips are the real trajectories of the GW differential, which form the Nakamura graph of the surface.
We can choose to label the real trajectories bounding the upper edge of each strip with the same integer that was assigned 
to the marked point lying on the upper edge of this strip. 
This gives us a labelling of the Nakamura graph associated to the surface, in which all the edges corresponding to the upper boundary of the same strip have the same label.
We call this labelling of a Nakamura graph the $S_d$ description, or the {\bf Hurwitz class} description, as the Nakamura graph
associated to this surface can be reconstructed from the Hurwitz class of the branched covering and vice versa.

\begin{figure}[t]
\begin{center}
\includegraphics[width=0.7\textwidth]{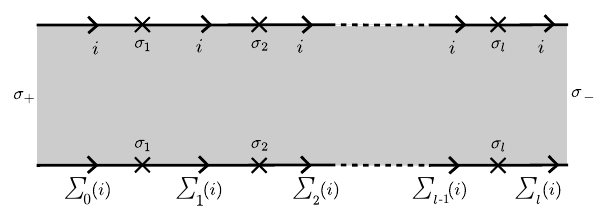}
\caption{The gluing of the strips can be read off from the Hurwitz tuple.}
\label{fig:strip}
\end{center}
\end{figure}

The labelling of the edges glued to the lower boundary of a strip are determined by the Hurwitz tuple.
On a strip in which the upper boundary is labelled by some integer $i\in\{1,2,\ldots, d\}$, the edge preceding the preimage of the first branch point is labelled by $\Sigma_0(i)$, where $\Sigma_0:=\sigma_+$. The edge proceeding the next branch point is labelled $\Sigma_1(i)$, where $\Sigma_1=\sigma_+ \sigma_1$; the next edge is labelled $\Sigma_2(i)$, with $\Sigma_2=\sigma_+\sigma_1\sigma_2$, and so on. This is shown in Figure \ref{fig:strip}.

\begin{figure}[ht]
\begin{center}
\includegraphics[width=0.7\textwidth]{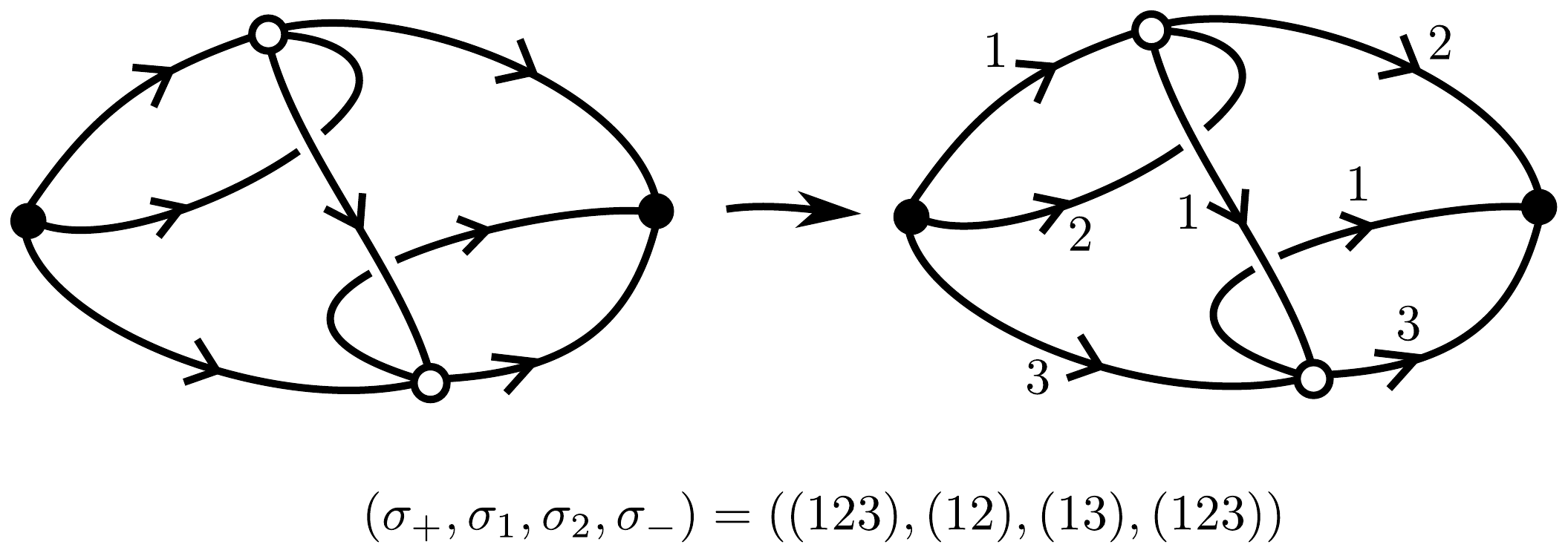}
\caption{A labelling of a Nakamura graph with an $S_d$ tuple.}
\label{fig:naktosd}
\end{center}
\end{figure}

Given a Nakamura graph associated to a surface, we can read off the Hurwitz tuple associated to a branched covering of the sphere as constructed above. The cyclic ordering of the edges at the incoming and outgoing poles correspond to $\sigma_+$ and $\sigma_-$ respectively, and the cyclic ordering of the incoming (or the outgoing) edges at the $i$th zero corresponds to $\sigma_i$. Each outgoing edge at a zero has the same label as the incoming edge located in the next clockwise position at the zero.
An example of a Nakamura graph with Hurwitz class labellings is given in Figure \ref{fig:naktosd} with the associated Hurwitz
tuple description.

Conversely, a Hurwitz tuple $(\sigma_+, \sigma_1, \ldots, \sigma_-)$ is enough to completely specify a Nakamura graph. 
As each Nakamura graph defines a cell in the light-cone cell decomposition of moduli space, we see that each Hurwitz class determines a cell in the LC cell decomposition.
In general, extra data is required to specify a particular point within this cell, as the permutation tuple alone does not encode the continuous data of the strip widths and the time coordinates of the zeroes.

One major advantage of the Hurwitz class description for Nakamura graphs is that there are only two conditions 
required for a permutation tuple to give a valid Nakamura graph. For a general $S_d$ tuple of $m+2$ permutations to describe a Nakamura graph:
\begin{itemize}
\item Each integer in $\{1,2,\ldots, d\}$ is permuted by at least one of ther permutations associated with the zeroes $\{\sigma_1,\sigma_2, \ldots, \sigma_m\}$. (This ensures that no trajectories connect poles directly to poles.)
\item The tuple $(\sigma_+, \sigma_1, \ldots, \sigma_m, \sigma_-)$ acts transitively on $\{1,2,\ldots, d\}$. (This ensures
that all associated Riemann surfaces are connected.)
\end{itemize}
All the other conditions given in Section \ref{sec:props} that a Nakamura graph must satisfy are guaranteed by the 
structure of the permutation tuple.

\begin{figure}[ht]
\begin{center}
\includegraphics[width=0.35\textwidth]{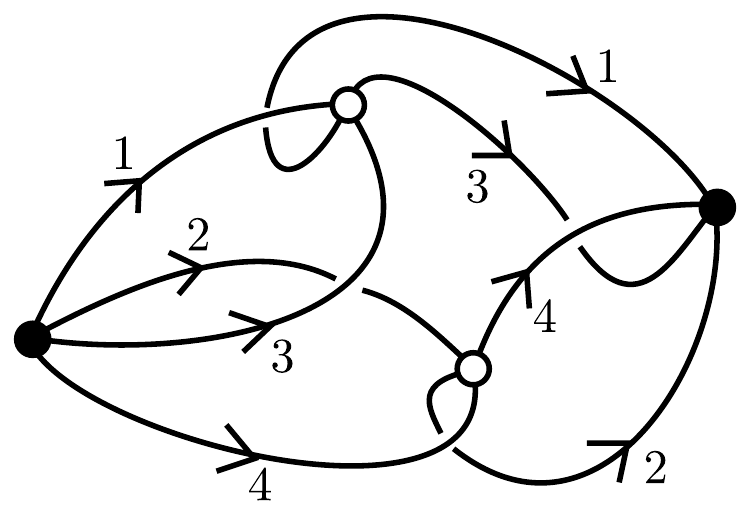} \qquad
\includegraphics[width=0.55\textwidth]{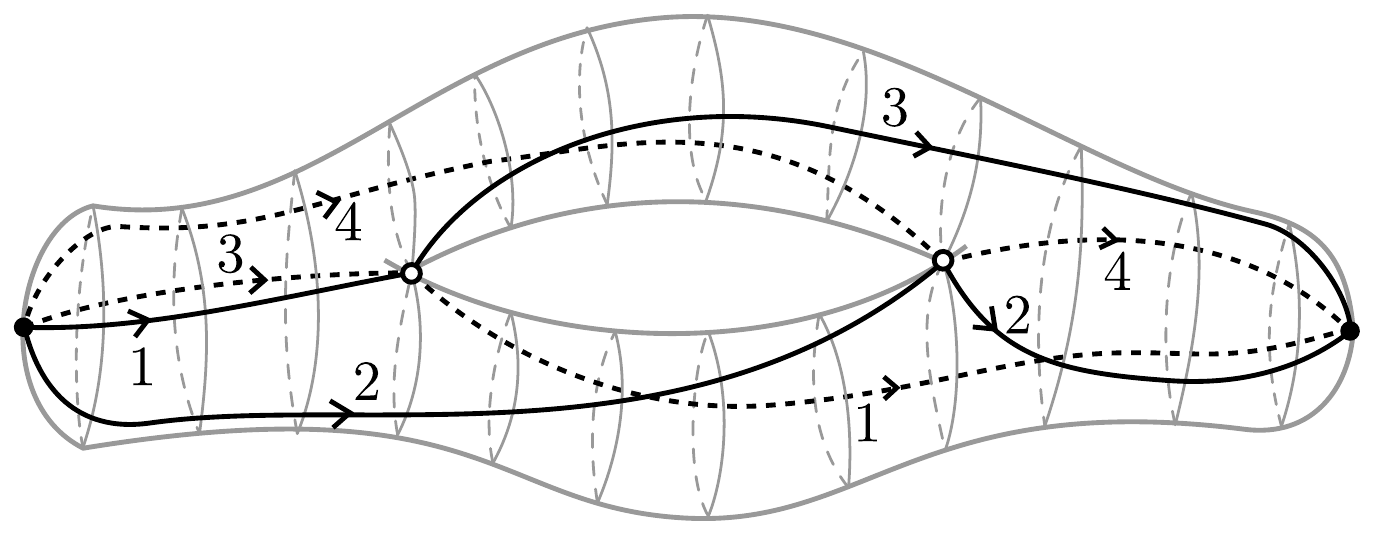}
\caption{A Nakamura graph in the $S_d$ picture, drawn embedded on the torus with the closed imaginary trajectories drawn in grey.}
\label{fig:aut4}
\end{center}
\end{figure}
\begin{figure}[ht]
\begin{center}
\includegraphics[width=0.35\textwidth]{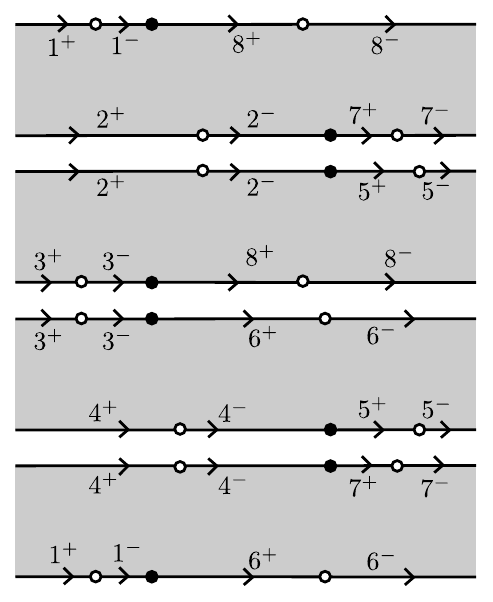} \qquad \includegraphics[width=0.35\textwidth]{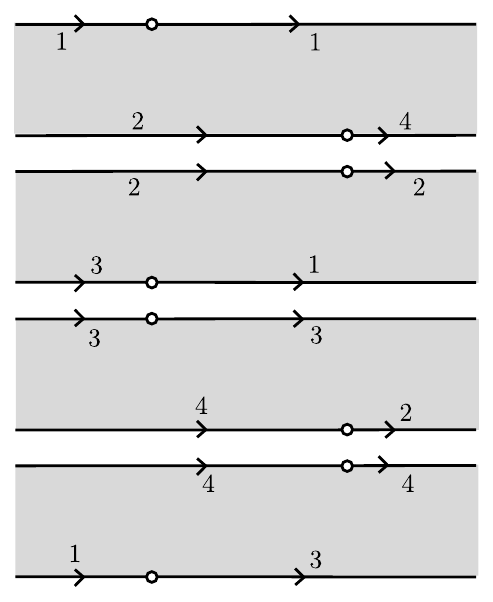}
\caption{The strip decomposition of the above graph in both the $S_{4d+2I}$ and the $S_d$ descriptions.}
\label{fig:strips}
\end{center}
\end{figure}

As an example of the strip decomposition of a surface via a Nakamura graph, and its description with an $S_d$ tuple, we consider again the example of a Nakamura graph with no internal lines and degree four, shown on the left of Figure \ref{fig:aut4}. This graph corresponds to a genus one surface with two marked points, drawn with the embedded Nakamura graph on the right of Figure \ref{fig:aut4}. 
This graph was described with $S_{4d+2I}$ and $S_{2d+2I}$ Belyi triples in \eref{eq:S4d2Itriple} and \eref{eq:S2d2Itriple}.
With the $S_d$ labellings, this graph has the associated Hurwitz class description
\bea
(\sigma_+, \sigma_1, \sigma_2, \sigma_-) = ((1234), (13), (24), (1234)).
\eea
The strip decomposition of the surface is shown in Figure \ref{fig:strips}, with the $S_{4d+2I}$ and
the $S_d$ labellings respectively. 
The cell associated to this graph in moduli space has real dimension $l+d-n=4$, which can be understood 
in terms of the continuous parameters of the strips. The residues of the poles are fixed to be $\pm r$.
There is an overall time translation symmetry of the strips, so we can set the first zero to have the time
coordinate $t=0$: the remaining zero has some time coordinate $t_1>0$.
We denote the widths of the strips with upper edges labelled $1,2,3,4$ by $b_1$, $b_2$, $b_3$, and $b_4$ respectively.
The sum of the widths of the strips is constrained to be $r$ due to the fact that the GW differential is simply $dz$ on each strip.
This gives four independent real parameters for the cell in moduli space, as required.

\begin{figure}[h!]
\begin{center}
\raisebox{10pt}{\includegraphics[width=0.35\textwidth]{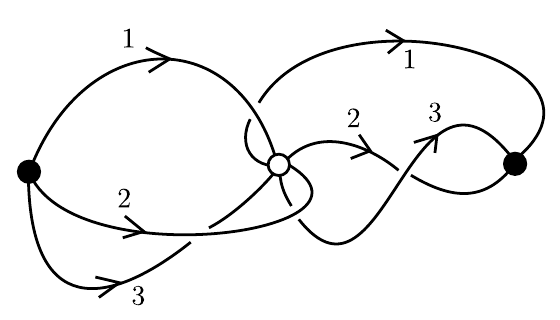}} \qquad
\includegraphics[width=0.50\textwidth]{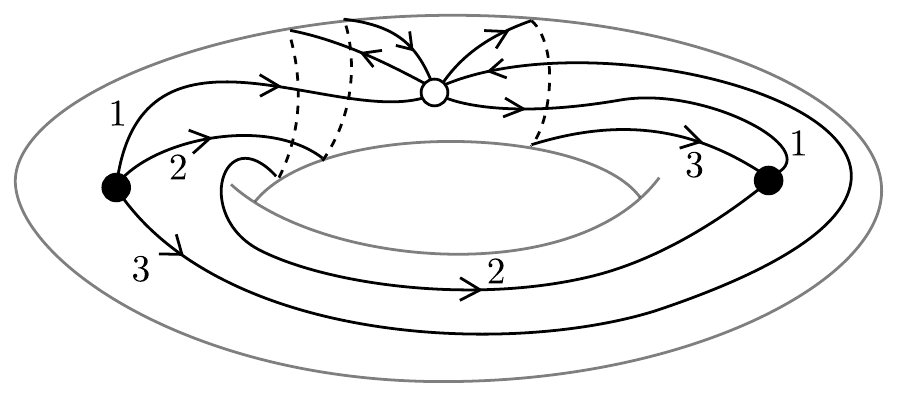}
\caption{A $\Delta=1$ Nakamura graph in the $S_d$ picture, and its embedding on the torus.}
\label{fig:Deltagraphs}
\end{center}
\end{figure}
In Figure \ref{fig:Deltagraphs} we have given another example of a Nakamura graph with $S_d$ labellings and its embedding on the torus. This graph has a non-zero branching constant $\Delta=1$, as the zero has a valency greater than four.
Its associated $S_d$ permutation tuple is
\bea
(\sigma_+, \tau_1, \sigma_-) = ( (123), (123), (123) ).
\eea

\subsection{Redundancies in the Hurwitz class description}

Given a Riemann surface with a Giddings-Wolpert differential, then there exists a unique branched covering of the sphere as constructed above up to equivalence, and so there exists a unique Hurwitz class associated to the surface. 
The cycles of the permutations in the Hurwitz class correspond to the vertices of the Nakamura graph. However,
there may be more than one Hurwitz class that can describe the same Nakamura graph. This is because a Hurwitz class
has a well-defined total ordering of the branch points, derived from the time coordinates of the zeroes, but a Nakamura 
graph generally only has a partial ordering on its zeroes derived from the orientation of the edges.

Consider the previous example of a Nakamura graph shown in Figure \ref{fig:aut4} and described
by the $S_d$ tuple
\bea
(\sigma_+, \sigma_1, \sigma_2, \sigma_-) = ((1234), (13), (24), (1234)). \label{eq:exchange}
\eea
The time coordinates of the zeroes associated to the permutations $(13)$ and $(24)$ satisfy $t_{(13)}< t_{(24)}$.
If we were to consider a surface with a different GW differential in which the time coordinates of the zeroes were interchanged
and $t_{(24)}<t_{(13)}$,
then the $S_d$ description of the graph would be 
\bea
(\sigma_+, \sigma_1, \sigma_2, \sigma_-) = ((1234), (24), (13), (1234)).
\eea
In addition, if we considered instead a surface where the time coordinates of the zeroes were identical, then the ramification of the 
branched cover of the sphere would no longer be simple, and the $S_d$ description of the graph would be
\bea
(\sigma_+, \sigma_1,  \sigma_-) = ((1234), (13)(24),  (1234)).
\eea
In all three of these cases, the Nakamura graph corresponding to the surface is {\it identical}.
A Nakamura graph only encodes an ordering on the time coordinates of the zeroes if there
is an oriented sequence of internal edges connecting the zeroes.

This redundancy makes the automorphisms of a Nakamura graph harder to determine in the Hurwitz class description
than in the Belyi triples descriptions. The set of permutations $\gamma\in S_d$ such that
\bea
(\gamma^{-1}\sigma_+\gamma,\ \gamma^{-1}\sigma_1\gamma,\ \ldots, \gamma^{-1}\sigma_m\gamma,\  \gamma^{-1}\sigma_-\gamma) = (\sigma_+, \sigma_1, \ldots, \sigma_m, \sigma_-)
\eea
are indeed automorphisms of the Nakamura graph, but they are not the only automorphisms. In some cases, there are 
permutations which map the $\sigma_i$ to each other
upon conjugation, which can preserve the structure of the associated Nakamura graph.
The example given above in Figure \ref{fig:aut4} has an automorphism group generated by the cycle
$\gamma=(1234)$, which interchanges the permutations $\sigma_1$ and $\sigma_2$ in the Hurwitz class
given in \eref{eq:exchange}.

To solve this redundancy in the Hurwitz class description, we introduce a new equivalence relation on the Hurwitz classes.
For a general tuple of $(m+2)$ permutations $(\sigma_+, \sigma_1, \ldots, \sigma_{m}, \sigma_-)$ describing a Nakamura graph arising from a branched covering of the sphere, each permutation $\sigma_i$ represents a set of zeroes with the same time coordinate. If there are two subsequent permutations $\sigma_i$ and $\sigma_{i+1}$
which are {\it disjoint} (the intersection of their moved-point sets is empty), then there are no internal
edges directly connecting any of the zeroes which correspond to the cycles in the permutations.
Any other branched covering with the $(m+1)$-permutation tuple 
\bea
(\sigma_+, \sigma_1, \ldots, \sigma_i \sigma_{i+1}, \ldots, \sigma_{m}, \sigma_-),
\eea
would have an identical Nakamura graph.

We define a binary relation on the set of permutation tuples by relating
\bea
(\sigma_+, \sigma_1, \ldots, \sigma_i, \sigma_{i+1}, \ldots, \sigma_{m}, \sigma_-)
\sim 
(\sigma_+, \sigma_1, \ldots, \sigma_i \sigma_{i+1}, \ldots, \sigma_{m}, \sigma_-).
\eea
whenever $\sigma_i$ and $\sigma_{i+1}$, $1\leq i <m$ are disjoint. This relation extends to an equivalence
relation on the set of tuples. The overall product of a tuple of permutations is unchanged by this relation,
and the overall action of conjugacy on tuples commutes with this relation, which means that this relation is 
a well-defined equivalence relation on the set of Hurwitz classes describing Nakamura graphs.
We call this relation {\bf slide-equivalence}, as it represents the ability to `slide' around the orderings of the zeroes
of a Nakamura graph when there are no internal edges connecting the zeroes. 
With this equivalence relation, each slide-equivalence class corresponds to a unique Nakamura graph.

\subsection{The `reduced tuple' \texorpdfstring{$S_d$}{S d} picture} \label{sec:reduced}

There is a one-to-one correspondence between the Nakamura graphs and the slide-equivalent Hurwitz classes.
Up to conjugacy equivalence, we can canonically choose a representative element for each slide-equivalence class, 
which we call the {\bf reduced tuple} description of a Nakamura graph, and denote by $(\sigma_+, \tau_1, \ldots, \tau_m, \sigma_-)$.
Each slide-equivalence class has exactly one Hurwitz class specified by a representative tuple $(\sigma_+, \tau_1, \ldots, \tau_m, \sigma_-)$ 
with the property that  \emph{every cycle in $\tau_{i+1}$ shares a moved point with $\tau_i$}, for each $i=1,2, \ldots, (m-1)$.
Intuitively, this is the $S_d$ tuple gained from taking a Nakamura graph and placing as many cycles as possible in the earliest permutation. Graphically, this tuple is gained by sliding the zeroes around so that as many zeroes as possible are 
vertically adjacent in the earliest position, and then subsequently as many zeroes as possible are arranged in the second earliest
position, and so on.

The reduced tuple has the property that the graph automorphisms do not exchange cycles between different $\tau_i$.
This means that the automorphisms of a Nakamura graph described by a reduced tuple are precisely those $\gamma\in S_d$ such that
\bea
(\gamma^{-1} \sigma_+ \gamma, \gamma^{-1} \tau_1 \gamma, \ldots, \gamma^{-1} \tau_m \gamma, \sigma_-)
= (\sigma_+, \tau_1, \ldots, \tau_m, \sigma_- ).
\eea

\begin{figure}[ht]
\begin{center}
\includegraphics[width=0.4\textwidth]{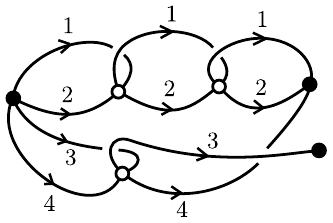}
\caption{A Nakamura graph in the $S_d$ picture.}
\label{fig:reduced}
\end{center}
\end{figure}
As an example, consider the slide-equivalence class describing the Nakamura graph given in Figure \ref{fig:reduced}.
With the labelling shown in the figure, this graph can be described by the tuples
\bea
(\sigma_+, \sigma_1, \sigma_2, \sigma_3, \sigma_-) &=& ((1234), (34), (12), (12), (142)), \\
(\sigma_+, \sigma_1, \sigma_2, \sigma_-) &=& ((1234), (12)(34), (12), (142)), \\
(\sigma_+, \sigma_1, \sigma_2, \sigma_3, \sigma_-) &=& ((1234),  (12), (34), (12), (142)), \\
(\sigma_+, \sigma_1, \sigma_2, \sigma_-) &=& ((1234), (12), (12)(34), (142)), \\
(\sigma_+, \sigma_1, \sigma_2, \sigma_3, \sigma_-) &=& ((1234), (12), (12), (34), (142)).
\eea
All these tuples lie in different Hurwitz classes, but their associated classes lie in the same slide-equivalence class.
This slide-equivalence is associated to the sliding the time coordinate of the zero associated to the transposition 
$(34)$. Of the five elements of the slide-equivalence class, the reduced tuple is
\bea
(\sigma_+, \tau_1, \tau_2, \sigma_-) &=& ((1234), (12)(34), (12), (142)),
\eea
as it is the only element which has the property that every cycle in $\tau_{i+1}$ shares a moved point with $\tau_i$ for all $i$.

\section{Counting of graphs with matrix models}\label{sec:hmm}

In this section we consider graphs with no internal edges and a single incoming pole, described by 
triples of permutations in the reduced $S_d$ description. 
Any such graph is described by a triple $(\sigma_+, \tau, \sigma_-)$, with $\sigma_+\tau\sigma_-=1$,
where $\sigma_+$ consists of a single $d$-cycle, $\tau$ consists of $l$ disjoint cycles corresponding to the 
internal vertices, and $\sigma_-$ consists of $(n-1)$ disjoint cycles corresponding to the outgoing poles.
There are correlators in the Gaussian and the complex matrix models that directly correspond to counting triples
of permutations multiplying to one. This allows us to apply known explicit expressions for matrix model
correlators to the counting of Nakamura graphs.

In Section 3, we stated that a Nakamura graph is associated to a cell $\cC$ in the LC cell decomposition of $\cM_{g,n}$ 
with real dimension
\bea 
\mbox{dim}_{\mathbb{R}}(\cC)=6g-6+2n- (2\Delta+I).
\eea
The genus of the graph is $g$, the number of poles is $n$, the number of internal edges connecting zeroes to zeroes is $I$, and the branching number $\Delta$ is defined in terms of the valencies of the $l$ zeroes by the formula 
\bea
\Delta = \sum_{i=1}^l \frac{1}{2}(v_i-4).
\eea
The degree $d$ is given in terms of $g$, $n$, $\Delta$, and $I$ by the formula
\bea
d = 2(2g-2+n) - (\Delta + I).
\eea
The total dimension of moduli space is $6g-6+2n$, and so the {\it codimension} of a cell in moduli space associated to a graph $\cG$ is $(2\Delta+I)$.

The top-dimensional cells of moduli space are associated to graphs with $\Delta=0$ and $I=0$. 
The zeroes of these graphs have valency four, and each zero can be described in the $S_d$ description
by a cycle permuting two labels (a transposition).
The permutation triples corresponding to graphs in the codimension zero cell are of the form $(\sigma_+, \tau, \sigma_-)$,
where $\tau$ is in $\cT = [2^l]$, the $S_d$ conjugacy class consisting of elements that are composed of $l=d/2$ disjoint 2-cycles.
For graphs with $I=0$ and $\Delta>0$, some of the zeroes will have valency greater than four, which correspond to 
cycles with size greater than two. For example, a graph with $\Delta=1$ is described by some $\tau$ in the conjugacy
class of elements with $(l-1)$ 2-cycles and one 3-cycle, $\cT = [2^{l-1},3]$.
A graph with $\Delta=2$ is described by some $\tau$ in either the conjugacy class 
$\cT_1 = [2^{l-2}, 3^2]$ or in $\cT_2 = [2^{l-1}, 4]$.

The counting of permutation triples where two permutations are in the classes $[d]$ and $[2^{d/2}]$ respectively 
is known to correspond to a correlator in the Gaussian matrix model. In Section \ref{sec:gaussian}, we use this
link to find the contribution to the orbifold Euler characteristic that comes from graphs in the top-dimensional cell. This can be 
checked explicitly against the tables derived in \cite{Nak}.
Also, the counting of permutation triples in more general classes is known to correspond to correlators in the complex 
matrix model. In Section \ref{sec:complex}, we can find the contributions to the orbifold Euler characteristic coming from graphs of higher codimension. This is checked against graphs counted directly by the software GAP.

\subsection{The Gaussian Hermitian matrix model}\label{sec:gaussian}
Triples of permutations of the form $(\sigma_+, \tau, \sigma_-)$, where $\sigma_+\tau\sigma_-=1$, $\sigma_+\in[2l]$ and $\tau \in [2^l]$, arise in the combinatorics of 
the Gaussian Hermitian matrix model. 
We can develop a link between the counting of top-dimensional graphs in moduli space with a single incoming pole and Gaussian matrix model correlators as follows.

First, note that a single-trace correlator in the Gaussian Hermitian matrix model can be written as
\bea 
\tr X^{d} = X^{i_1}_{ i_{ \sigma (1) } }  \cdots X^{i_d }_{ i_{\sigma(d) } } 
\eea
with $\sigma = ( 1, 2, \cdots , d )$. In other words, when we have a single trace, the lower indices are a cyclic permutation of the upper indices. 
Now when we perform the Wick contraction on the correlator, we are summing over pairings of $d$ objects,
e.g. $ (1,2) ( 3,4) \cdots ( d-1, d)$. Each pairing corresponds to a permutation $\tau$ in the class 
$[2^l]$ where $l = d/2$. The Matrix model correlator of a single trace $\tr X^d$ can be written in terms of these two permutations 
rather simply:
\bea 
\cor{ tr X^d } =  \sum_{ \tau \in [2^l] } \sum_{ \alpha \in S_d } \delta ( \sigma \tau \alpha ) N^{ C_{\alpha } },
\eea
where the delta function imposes the condition that the three permutations multiply to $1$,
and $C_{\alpha } $ is  the number of cycles in the product $ \alpha = (\sigma \tau)^{-1}$.
We can also introduce a sum over the conjugacy class of single-cycles of length $d$ accompanied by a factor of $|[d]|=(d-1)!$ 
without changing value of the correlator:
\bea 
\cor{ tr X^d } =  \frac{1}{(d-1)!}\sum_{\sigma \in [d]} \sum_{ \tau \in [2^l] } \sum_{ \alpha \in S_d } \delta ( \sigma \tau \alpha ) N^{ C_{\alpha } }.
\eea

Now, consider the equivalence classes of triples $(\sigma, \tau, \alpha)$, where
\bea 
( \sigma' , \tau' , \alpha' ) \sim ( \gamma \sigma \gamma^{-1} , \gamma \tau \gamma^{-1} , \gamma \alpha \gamma^{-1} ).
\eea
These equivalence classes correspond precisely to the Nakamura graphs with a single incoming pole, no internal edges, and $\Delta=0$
 in the $S_d$ description. The permutation $\sigma$ corresponds to the single incoming pole,
$\tau$ corresponds to the zeroes, and $\alpha$ to the outgoing poles. The number of poles in a Nakamura graph given by such a tuple is $ C_{\sigma} + C_{\alpha }$, which
is equal to $n$. As $C_{\sigma}=1$, and we are interested in graphs corresponding to surfaces with $n$ marked points, we can
consider just the permutation tuples with $C_{\alpha}=n-1$, and so consider the coefficient of $N^{n-1}$ in the correlator:
\bea
\text{Coefficient}(\cor{ tr X^d }, N^{n-1}) = \frac{1}{(d-1)!}\sum_{\sigma \in [d]} \sum_{ \tau \in [2^l] } \sum_{\substack{ \alpha\in S_d\\ C_\alpha=(n-1)}} \delta ( \sigma \tau \alpha ).
\eea
We can split the sum over $\alpha$ into a sum over distinct conjugacy classes $[\hat{\alpha}]$, each consisting of $(n-1)$ cycles, and a sum over each individual class with $(n-1)$ cycles $[\hat{\alpha}]$,
\bea
\text{Coefficient}(\cor{ tr X^d }, N^{n-1}) = \frac{1}{(d-1)!}\sum_{\substack{\text{classes}\\ [\hat{\alpha}]}} \sum_{\sigma \in [d]} \sum_{ \tau \in [2^l] } \sum_{\alpha\in [\hat{\alpha}]} \delta ( \sigma \tau \alpha ).
\eea
Now the sum 
\bea 
\frac{ 1}{  d! } 
 \sum_{ \sigma \in [d ] }  \sum_{ \tau \in [2^l] } \sum_{\alpha\in [\hat{\alpha}]}  \delta ( \sigma \tau \alpha ) 
\eea
can be written in terms of equivalence classes of permutation triples. 
By the Orbit-Stabiliser theorem, the number of times each equivalence class appears in the sum is
\bea 
\frac{d!}{|\text{Aut}(\{\sigma,\tau,\alpha\})| }   
\eea
where Aut$(\{\sigma,\tau,\alpha\})$ is the order of the automorphism group of the triple. Each equivalence class corresponds
to a distinct bipartite graph. This means that 
\bea 
&& \frac{ 1}{d! } \sum_{\substack{\text{classes}\\ [\hat{\alpha}]}} \sum_{ \sigma \in [d ] }  \sum_{ \tau \in [2^l] } \sum_{ \alpha \in  [\hat{\alpha} ] } \delta ( \sigma \tau \alpha ) \cr
&& = \sum_{\substack{\text{classes}\\ [\hat{\alpha}]}} \sum_{  \substack{ \text{equiv classes}\\ \text{of triples}}  }  \frac{ 1 }{|\text{Aut}(\sigma,\tau,\alpha)|  }  \cr 
&& = \sum_{ \text{graphs} } \frac{ 1 }{|\text{Aut}(\sigma,\tau,\alpha)| }. \label{eq:eccontrib}
\eea
This sum is taken over all the graphs specified by a permutation triple $(\sigma, \tau, \alpha)$ with one incoming pole and $n-1$ outgoing poles.
This is exactly the sum that Nakamura performed to find the contribution of the 
top-dimensional cells to the orbifold Euler characteristic of $\cM_{ g, n }$. 
We conclude that the contribution of the top cell of $\cM_{g,n}$ to the orbifold Euler characteristic is
\bea 
\chi_{\text{top}}(g,n) = \frac{1}{d}\times \text{Coefficient}(\cor{ tr X^d }, N^{n-1}).
\eea

There is a generating function for correlators of single traces in the Gaussian Hermitian matrix model, due
to Harer and Zagier:
\bea 
C ( x , N ) &=&
\sum_{ l=1 }^\infty \cor{\tr (  X^{2l } ) } \frac{  x^{2l} }{ ( 2l  -1) !! } \\ &=& \frac{ 1 }{ 2 x^2 } \left (   \left ( \frac{  1 + x^2 }{ 1 - x^2   } \right )^{ N }  -1 \right).
\eea
This means that the contribution to the top-dimensional cell can be read off:
\bea 
\chi_{\text{top}}(g,n) = \frac{(d-1)!!}{d}\  \text{Coefficient}( C(x,N), x^dN^{n-1}).
\eea
We can calculate exactly the coefficient of $N^{n-1}$ in this expression.
Noting that
\bea
C(x,N) = \frac{1}{2x^2}\left[\exp\left(N \log \left(\frac{1+x^2}{1-x^2}\right)\right) - 1\right],
\eea
we differentiate this $(n-1)$ times with respect to $N$ to see that
\bea
\hbox{ Coefficient } ( C(x,N), N^{n-1} ) = \frac{1}{2x^2(n-1)!}\left[\log \left(\frac{1+x^2}{1-x^2}\right) \right]^{n-1}.
\eea
The contribution to the Euler characteristic is therefore
\bea
\chi_{\text{top}}(g, n) &=& \frac{(d-1)!!}{2d(n-1)!}
\hbox{ Coefficient } \left( \frac{1}{x^2}\left[ \log \left(\frac{1+x^2}{1-x^2}\right) \right]^{n-1}, \quad x^d \right) \\
&=& \frac{(d-1)!!}{2d(n-1)!}
\hbox{ Coefficient } \left( \left[\log \left(\frac{1+w}{1-w}\right)\right]^{n-1}, \quad w^{(n-1) +2g}\right),
\eea
where $d= 2(2g-2+n)$, and we have substituted $w=x^2$ in the final equation. Written purely in terms of $g$ and $n$, the expression for the Euler characteristic contribution is
\bea
\chi_{\text{top}}(g, n) = \frac{(4g-5+2n)!}{2^{2g-3+n}(n-1)!(2g-2+n)!} \hbox{ Coefficient } \left( \log \left(\frac{1+w}{1-w}\right)^{n-1},  w^{(n-1) +2g}\right). \qquad
\eea
This expression matches the values found by counting graphs in Nakamura's paper.

In the case $n=2$, the series expansion of the generating function can be found exactly. We have
\bea
\log\left(\frac{1+w}{1-w}\right)=2\sum_{g=0}^\infty \frac{w^{2g+1}}{(2g+1)},
\eea
so we deduce that
\bea
\chi_{\text{top}}(g, 2) = \frac{ ( 4g )! }{ 2^{2g} ( 2g ) ! }  \frac{ 1 }{ 4 g } \frac{ 1 }{ 2g+1 } = \frac{ ( 4g -1 ) ! }{2^{ 2g } ( 2g +1)! } \label{chig2}.
\eea
This sequence, starting at $g=1$ is : 
\bea\label{matrixsequence} 
\frac{ 1 }{ 4 } , \frac{ 21 }{ 8} , \frac{ 495 }{ 4 } , \frac{ 225225 }{16 } \cdots 
\eea
The first three terms in this sequence correspond to the tables of data in Nakamura. 
The case $ ( g , n ) = ( 4,2 )$ was not provided in Nakamura, so the value $\frac{ 225225 }{16 }$ 
is a prediction, as are the infinite series of coefficients \eref{chig2}.
However, the top-cell contribution in the $(g,n)=(4,2)$ case was confirmed directly by counting the graphs using the software GAP.

\subsection{The complex matrix model}\label{sec:complex}

Let $\cT$ be the conjugacy class of $S_d$ elements $[2^{k_2}3^{k_3}\ldots d^{k_d}]$. 
Choose a representative element $\hat{\sigma}_+ \in [d]$ and $\hat{\tau}\in \cT$.
The complex matrix model correlator of a holomorphic trace and an antiholomorphic product of traces corresponding
to these classes is
\bea
\cor{\tr (\hat{\sigma}_+ Z^{\otimes d})\tr (\hat{\tau}Z^{\dagger \otimes d})} &:=&
\cor{\tr Z^d (\tr Z^{\dagger 2})^{k_2}(\tr Z^{\dagger 3})^{k_3}\ldots (\tr Z^{\dagger d})^{k_d}} \\
&=& \frac{d}{|\cT|}\sum_{\sigma_+\in [d]}\sum_{\tau \in \cT}\sum_{\sigma_- \in S_d}N^{C_{\sigma_-}}\delta(\sigma_+\tau\sigma_-).
\eea
As in the Hermitian matrix model, this correlator is a sum over conjugacy classes of permutation triples that
multiply to one. Splitting up the sum over $\sigma_-\in S_d$, we can write
\bea
\cor{\tr (\hat{\sigma}_+ Z^{\otimes d})\tr (\hat{\tau}Z^{\dagger \otimes d})} 
= \frac{d}{|\cT|}\sum_{n=2}^{d-1} N^{n-1} \sum_{\sigma_+\in [d]}\sum_{\tau \in \cT}\sum_{\substack{\sigma_-\in S_d\\ C_{\sigma_-}=n-1}} \delta(\sigma_+\tau\sigma_-).
\eea
This expression is a sum over the Nakamura graphs with $n$ external points and internal vertex structure given 
by $\cT$:
\bea
\cor{\tr (\hat{\sigma}_+ Z^{\otimes d})\tr (\hat{\tau}Z^{\dagger \otimes d})}  = \frac{d! d}{|\cT|}
\sum_{n=2}^{d-1} N^{n-1} \sum_{\cG}\frac{1}{|\text{Aut}(\cG)|}.
\eea
The sum over $\cG$ is taken over all graphs with $(n-1)$ outgoing poles with internal structure given by $\cT$.
This sum appears in the orbifold Euler characteristic of moduli space of genus $g$ with $n$ marked points:
defining the contribution to the orbifold Euler characteristic coming from a class $\cT$ by the formula
\bea
\chi_\cT (g, n) = \sum_{\cG} \frac{1}{\text{Aut}(\cG)}, \label{eq:classcontrib}
\eea
we can state that the contribution to the Euler characteristic coming from graphs with class $\cT$ is
\bea
\chi_\cT(g,n) = \frac{|\cT|}{d!d}\ \text{Coefficient}(\cor{\tr (\hat{\sigma}_+ Z^{\otimes d})\tr (\hat{\tau}Z^{\dagger \otimes d})}, N^{n-1}).
\eea
It is useful to recall that the parameters $ k_i $ defining $ \cT $ relate to the parameters in section \ref{sec:review}
by 
\bea
&& l  = \sum_{ i =2}^d k_i \cr 
&& \Delta = \sum_{ i =2}^d ( i - 2 ) k_i  = d - 2 l 
\eea

The complex matrix model correlator can be calculated by using character sums. In \cite{GRW1403}, it is shown that 
\bea
\cor{\tr Z^d (\tr Z^{\dagger 2})^{k_2}(\tr Z^{\dagger 3})^{k_3}\ldots (\tr Z^{\dagger d})^{k_d}} =
d!\sum_{t=0}^d \sum_{\substack{S \subset \{1,2,\ldots l \} \\ |S|=t}}(-)^{l-t}\binom{N+\sum_{i\in S}k_i}{d+1}. \nn
\eea
\bea
= d! \sum_{r_1=0}^{k_1}\sum_{r_2=0}^{k_2}\ldots \sum_{r_d=0}^{k_d} (-)^{k_1+\ldots +k_d - r_1 - \ldots - r_d}
\binom{k_1}{r_1}\ldots\binom{k_d}{r_d}
\binom{N+\sum_{j=1}^d jr_j}{d+1}.
\eea
The size of the conjugacy class $\cT$ is 
\bea
|\cT| = \frac{d!}{k_2!2^{k_2}k_3!3^{k_3}\ldots k_d! d^{k_d}}.
\eea
This gives us an explicit expression for the orbifold Euler characteristic contribution from the class $\cT = [2^{k_2}3^{k_3}\ldots d^{k_d}]$:
\begin{multline}
\chi_\cT(g,n) =
\frac{(d-1)!}{k_2!2^{k_2}k_3!3^{k_3}\ldots k_d! d^{k_d}}
\sum_{r_1=0}^{k_1}\ldots \sum_{r_d=0}^{k_d}
(-)^{k_1+\ldots +k_d - r_1 - \ldots - r_d}
\binom{k_1}{r_1}\ldots\binom{k_d}{r_d}
\times \\ \times
\text{Coefficient}\left[  
\binom{N+\sum_{j=1}^d jr_j}{d+1} 
, N^{n-1}\right].
\end{multline}

This formula can reproduce the Euler characteristic contributions for cells of codimension zero.
For fixed $g$, $n$ with $\Delta=0$, then the degree $d$ is $2(2g-2+n)$, the number of zeroes is
$l=d/2 = 2g+n-2$, and the contribution to the Euler characteristic is
\begin{multline}
\chi_{[2^l]}(g,n) = \frac{(2l-1)!}{l!2^l}\sum_{r_2=0}^{l}(-)^{l-r_2}\binom{l}{r_2}\text{Coefficient}\left[
\binom{N+2r_2}{2l+1}
, N^{n-1}\right] \\
= \frac{(4g+2n-5)!}{(2g+n-2)!2^{2g+n-2}}
\sum_{r_2=0}^{2g+n-2}(-)^{2g+n-2-r_2}\binom{2g+n-2}{r_2}
\text{Coefficient}\left[
\binom{N+2r_2}{4g+2n-3}
, N^{n-1}\right]. \label{eq:matches}
\end{multline}
This formula has been checked numerically for graphs of degree $d\leq 9$ against the tables in Nakamura.
We have quoted the relevant top-cell graphs in Table \ref{tab1}, using the notation $[a]\times n$ to denote
$n$ graphs with cyclic automorphism groups of order $a$.
The contribution to the Euler character calculated by counting the graphs and using the formula \eref{eq:eccontrib} 
exactly matches the results derived from \eref{eq:matches}.

\begin{table}
\begin{center}
{\renewcommand{\arraystretch}{1.25}
  \begin{tabular}{| c | c | c |}
    \hline
    $(g,n)$ & $\chi_{[2^l]}(g,n)$ & $\Delta=0$ Graphs \\ \hline
    $(0,5)$ & $\tfrac{5}{6}$ & $[2]\times 1, [3]\times 1$ \\ \hline
    $(0,6)$ & $\tfrac{7}{4}$ & $[1]\times 1, [2]\times 1, [4]\times 1$ \\ \hline
    $(0,7)$ & $\frac{21}{5}$ & $[1]\times 3, [2]\times 2, [5]\times 1$ \\ \hline
    $(1,3)$ & $\tfrac{5}{3}$ & $[1]\times 1, [2]\times 1, [6]\times 1$ \\ \hline
    $(1,4)$ & $\tfrac{35}{4}$ & $[1]\times 7, [2]\times 3, [4]\times 1$ \\ \hline
    $(1,5)$ & 42 & $[1]\times 38, [2]\times 8$ \\ \hline
    $(2,2)$ & $\tfrac{21}{8}$ & $[1]\times 2, [2]\times 1, [8]\times 1$ \\ \hline
  \end{tabular}
} \qquad
{\renewcommand{\arraystretch}{1.25}
  \begin{tabular}{| c | c | c |}
    \hline
    $(g,n)$ & $\chi_{[2^{l-1}3]}(g,n)$ & $\Delta=1$ Graphs \\ \hline
    $(0,5)$ & 1 & $[1]\times 1$ \\ \hline
    $(0,6)$ & 3 & $[1]\times 3$ \\ \hline
    $(0,7)$ & $\tfrac{28}{3}$ & $[1]\times 9, [3]\times 1$ \\ \hline
    $(1,3)$ & 3 & $[1]\times 3$ \\ \hline
    $(1,4)$ & 20 & $[1]\times 20$ \\ \hline
    $(1,5)$ & $\tfrac{350}{3}$ & $[1]\times 116, [3]\times 2$ \\ \hline
    $(2,2)$ & 7 & $[1]\times 7$ \\ \hline
  \end{tabular}
}
\end{center}
\caption{The number of graphs and their automorphism group sizes against $\chi_{[2^l]}(g,n)$ and $\chi_{[2^{l-1}3]}(g,n)$ for different values of $g$ and $n$. The notation $[a]\times n$ denotes $n$ graphs with cyclic automorphism group of order $a$.}\label{tab1}
\end{table}
For graphs with $\Delta=1$, the conjugacy class $\cT$ is of the form $\cT=[2^{l-1},3]$ for some $l$. We have
$d= 3 + 2(l-1) = 2(2g-2+n) -1$, so $l=2g+n-3$. 
The Euler characteristic sum is 
\begin{multline}
\chi_{[2^l]}(g,n) = \frac{2l!}{(l-1)!2^{l-1}3}\sum_{r_2=0}^{l-1}\sum_{r_3=0}^{1}(-)^{l-r_2-r_3}\binom{l-1}{r_2}\binom{1}{r_3}\text{Coefficient}\left[
\binom{N+2r_2+3r_3}{2l+2}
, N^{n-1}\right] \\
= \frac{(4g+2n-6)!}{(2g+n-4)!2^{2g+n-4}3}
\sum_{r_2=0}^{2g+n-4}\sum_{r_3=0}^1(-)^{2g+n-3 -r_2-r_3 }\binom{2g+n-4}{r_2}\binom{1}{r_3} \\
\text{Coefficient}\left[
\binom{N+2r_2+3r_3}{4g+2n-4}
, N^{n-1}\right].\label{eq:coeff}
\end{multline}
A program was written in GAP to count all the graphs with $I=0$ and $\Delta=1$ for a given genus
$g$ and number of external points $n$; the results are tallied in Table \ref{tab1}. The formula \eref{eq:coeff} precisely
matches the calculation of the contribution to the Euler character produced by using the explicit graph counting and
\eref{eq:classcontrib}.

\section{Counting Nakamura graphs in the \texorpdfstring{$S_d$}{S d} picture using GAP}\label{sec:gap}

Nakamura was able to confirm that the graphs gave a valid cell decomposition of moduli space by finding
all the graphs in a given moduli space $\cM_{g,n}$, and showing that the orbifold Euler characteristic
\bea
\chi(g, n) = \sum_{\cG}(-1)^\text{dim}\frac{1}{|\text{Aut}(\cG )|}
\eea
matches the orbifold Euler characteristic predicted by Harer and Zagier. 
In the above expression, the sum is taken over all inequivalent graphs $\cG$, each with
automorphism group Aut$(\cG)$, and `dim' is the dimension of the cell in moduli space associated to each graph.

Each Nakamura graph corresponds to a slide-equivalence class of Hurwitz classes, in each of which there is a unique
Hurwitz class of reduced tuples. In this section, we describe how we can use the reduced $S_d$ tuple description of graphs to 
count Nakamura graphs algorithmically.
We were able to implement this algorithm with the software GAP to reproduce the tables of Nakamura's paper, in which the 
graphs with a single incoming pole were counted with their automorphism groups. 
For computational efficiency, the algorithm works by taking as input a maximum value of the degree $d$ of the permutation 
groups $S_d$, and fixing the incoming pole to be of the form $(1,2,\ldots, d)$. The algorithm then considers in turn every 
conjugacy class of the reduced permutations $\tau_1,\ldots, \tau_m$ that can yield a Nakamura graph, and calculates the 
automorphisms of the allowed graphs. The permutation $\sigma_-$
is determined by $\sigma_- = (\sigma_+\tau_1\ldots\tau_m)^{-1}$, and the number of cycles in this permutation gives the number of outgoing poles $n-1$.

The algorithm proceeds as follows:
\bi
\item 
First, fix a value of the (graph) Euler characteristic $\chi=2-2g-n$. From \ref{defdmax}, this gives the maximum number of faces of the associated
Nakamura graphs. It also gives an upper bound on the degree $d_{max}=2|\chi|$ of the permutation groups $S_d$ that can describe graphs of this Euler characteristic.
\item Allow $l$ to scroll over the range \eref{rangel}, $1 \leq l \leq (-\chi) $.
For each $l$, \eref{defDelta} gives us $\Delta$.

\item Given $l$ and $\Delta$, find all the possible valencies of the internal vertices.
These valencies can be described by an {\it unordered} tuple of $S_d$ conjugacy classes $[[\sigma_1],\ldots, [\sigma_l]]$.
Each conjugacy class is of the form $[i]$ for some $i_k>1$; that is, each permutation in the conjugacy
class consists of a cycle of length $i$ and $(d-i)$ cycles of length one. We call these the `unreduced class tuples'.

In the $S_d$ picture of describing Nakamura graphs, a zero with valency $2k$ is described by a $k$-cycle.
The branching number $\Delta$ is related to the zeroes connecting to more than four edges.
More precisely, the possible valencies of the vertices correspond to the possible ways of
partitioning $\Delta+2l$ indistinguishable objects into $l$ sets with at least 2 elements.
(For example, if we had $d_{max}=10$, $L=3$, and $\Delta=2$, then the only possible unreduced class tuples
are $[[3],[3],[2]]$ and $[[4],[2],[2]]$.)
\item For each unreduced class tuple, find all the possible `reduced class tuples'. 
A reduced class tuple is an ordered list of $S_d$ conjugacy classes $(\cT_1, \cT_2, \ldots, \cT_m)$, where $m\leq l$, formed by merging together the classes from an unreduced class tuple in some way.
Each $\cT_i$ is of the form $[a_1, a_2, \ldots, a_{k_i}]$, where the $a_i$ are lengths of cycles
from the unreduced class tuple which are greater than one. Each cycle length from the unreduced
class tuple appears in exactly one $\cT_i$. 

For example, the unreduced tuple $([3],[3],[2])$ can be combined as $\cT_1=[3,3,2]$ with $m=1$, or
as $(\cT_1, \cT_2) \in \{([3,3],[2]), ([2], [3,3]), ([3,2],[3]), ([3],[3,2])\}$ for $m=2$,
or as $(\cT_1, \cT_2, \cT_3) \in \{ ([3],[3],[2]), ([3],[2],[3]), ([2],[3],[3])\}$ for $m=3$.

We are only interested in the reduced class tuples which can give valid Nakamura graphs
in the reduced tuple picture. This means we should discard any sequence of class tuples
in which there is some $i\in\{1,\ldots m-1\}$ such that $\tau_i\in \cT_i$ permutes fewer integers 
than the number of disjoint cycles in $\tau_{i+1}\in \cT_{i+1}$.

As another example, we could partition the unreduced tuple $([2],[2],[2],[2])$ into reduced tuples with
$m=1,2,3,4$. The only possible $m=1$ reduced tuple is $\cT_1=[2,2,2,2]$; the $m=2$ reduced tuples
are  $(\cT_1,\cT_2) \in \{ ([2,2,2],[2]) ([2,2],[2,2]) \}$; the $m=3$ reduced tuples are
$(\cT_1,\cT_2,\cT_3) \in \{ ([2,2],[2],[2]), ([2],[2,2],[2]), ([2],[2],[2,2]) \}$, and the 
only $m=4$ reduced tuple is $(\cT_1,\cT_2,\cT_3,\cT_4)=([2],[2],[2],[2]).$
Note that $(\cT_1,\cT_2) = ([2], [2,2,2])$ is \emph{not} an allowed reduced tuple: 
a permutation $\tau_1\in \cT_1$ moves two points while all permutations in $\cT_2$ have three non-trivial cycles,
so there are no permutations with this structure that can give a valid Nakamura graph in the reduced tuple description.

\item For each reduced class tuple, scroll over all the tuples $(\tau_1, \ldots, \tau_m)$ in the conjugacy classes 
$(\cT_1,\ldots,\cT_m)$. Keep the tuples with the following two properties:
	\bi
	\item For all $i\in\{1,\ldots m-1\}$, there is no cycle in $\tau_{i+1}$ that is disjoint from all cycles in $\tau_i$.
	\item The set of points moved by at least one of the $\tau_i$ is exactly $\{1,2,\ldots, d\}$ for some $d$.
	\ei
The value $d$ is the {\it degree} of the Nakamura graph associated to the tuple.
\item Act on the set of $\tau$-tuples with the same degree and same reduced class tuple with the group $\la (1,2,\ldots d)\ra$.
Each conjugacy class, together with $\sigma_+=(1,2,\ldots d)$,
gives a distinct Nakamura graph. Each graph has a cyclic automorphism group generated by $(1,2,\ldots,d)^k$,
where $k$ is the size of the conjugacy class of the $\tau$-tuple, and the size of the automorphism group is $d/k$.
\item Collate the graphs by genus $g$, the number of poles $n$, and the dimension of its cell in moduli space.
The number of disjoint cycles in $\sigma_- = (\sigma_+ \tau_1\ldots\tau_m)^{-1}$ is equal to $n-1$, 
the number of outgoing poles of the graph. The graph has genus $g$, where
\bea
g = -\frac{1}{2}\chi -\frac{n}{2}+1.
\eea
The dimension of the cell in moduli space associated to the graph is $l+d-n$.
\ei

This procedure can quickly generate all Nakamura graphs for $d_{max}\leq 10$ or so, and is capable of 
generating all Nakamura graphs for $d_{max}=12$, given sufficient time. However, the step of scrolling over
all tuples in $(\cT_1, \ldots, \cT_m)$ is very resource-intense, as a relatively small percentage of the
trial tuples give a valid Nakamura graph. (For $d=10$, about $6\%$ of trial tuples satisfy the two properties given above.)
In addition, the vast majority of Nakamura graphs have trivial automorphism group, so there is virtually a $d$-fold degeneracy 
in the graphs counted. 
For these reasons, we introduce in the next section a new structure within the reduced $S_d$ tuple description
that circumvents both these issues and results in a much more powerful method of counting Nakamura graphs.

\subsection{\texorpdfstring{$I$}{I}-structures } 

A Nakamura graph has $I$ internal edges that connect zeroes to zeroes. In the reduced $S_d$ tuple picture,
these edges are labelled by precisely those integers in $\{1,2,\ldots d\}$ which are permuted by more than one of
the $\tau_i$ in the tuple $(\tau_1, \tau_2,\ldots \tau_m)$. The integers which
are permuted by exactly one $\tau_i$ correspond to the external edges, which connect zeroes only to poles.
We can describe the structure of the
internal edges of the graph by creating a diagram that shows which permuted points are shared 
between the different $\tau_i$, which we call an \emph{$I$-structure.}

\begin{figure}[ht]
\begin{center}
\includegraphics[width=0.2\textwidth]{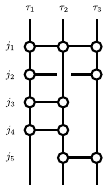}
\caption{An $I$-structure with five $I$-rows and three columns.}
\label{fig:Istr}
\end{center}
\end{figure}

An $I$-structure is a diagram consisting of $m$ parallel vertical edges, which we call `columns',
and several rows of horizontal edges, which we call `$I$-rows'.
Each $I$-row is a connected line of horizontal edges and vertices, with the vertices connecting
columns and horizontal edges.
An $I$-structure may contain the same $I$-row multiple times, and the 
$I$-rows of an $I$-structure are taken to be interchangeable.
All pairs of adjacent columns are connected by at least one edge of an $I$-row.
An example of an $I$-structure is given in Figure \ref{fig:Istr}.

There is a unique $I$-structure corresponding to each reduced tuple of permutations $\tau_i$, which represents
the internal edges of the associated Nakamura graph.
The $m$ columns correspond to the $m$ permutations in the tuple $(\tau_1, \ldots, \tau_m)$.
From the definition of a Nakamura graph, each integer in the set $\{1,2,\ldots d\}$ is permuted by at least one of the $\tau_i$.
If an integer $j$ is permuted by two or more of the $\tau_i$, then there is an $I$-row associated to this integer.
The vertices of this $I$-row are drawn on the columns corresponding to the $\tau_i$ which permute the integer $j$.
There is a horizontal edge associated to every consecutive pair of vertices along the $I$-row; 
these edges correspond to the internal edges of the Nakamura graph. Each vertex of the $I$-structure corresponds
to a zero (internal vertex) of the Nakamura graph, but there will in general be zeroes which do not correspond
to vertices of the $I$-structure.

The $I$-structure constructed from a permutation tuple is unique, but there will be many different
permutation tuples that have the same $I$-structure. For example, the $I$-structure given in Figure \ref{fig:Istr}
could be generated by the tuple of $S_6$ permutations
\bea
\tau_1=(1,2)(3,4), \qquad \tau_2=(1,3)(4,5), \qquad \tau_3=(1,2)(5,6).
\eea
The integers $\{1,2,3,4,5\}$ correspond to internal edges, and the integer `$6$' corresponds to external edges. 
If we conjugate the above tuple by some $\gamma\in S_6$, then we have the new tuple
\bea
(\tilde{\tau_1}, \tilde{\tau_2}, \tilde{\tau_3}) = (\gamma\tau_1\gamma^{-1}, \gamma\tau_2\gamma^{-1}, \gamma\tau_3\gamma^{-1}),
\eea
which is just a relabelling of the $\tau_i$ and so has the same $I$-structure.
In general, conjugate permutation tuples have the same $I$-structure, but there can also be distinct tuples which are not
conjugate which have the same $I$-structure.
An example of a permutation tuple that also generates the $I$-structure in Figure \ref{fig:Istr} and is not conjugate to the above
tuple is 
\bea
\tau_1=(1,2,3,4), \qquad \tau_2=(1,3,4,5), \qquad \tau_3=(1,2,5).
\eea 

\subsection{\texorpdfstring{$I$}{I}-structures for small \texorpdfstring{$I$}{I}}

For small values of $I$, we can explicitly list all the possible $I$-structures.
We start by considering $I=0$. Any graph with no internal edges must have $m=1$ in the reduced $S_d$ description, 
and so the tuples of these graphs take the form
\bea 
\sigma_+  \tau \sigma_- =1.
\eea
The Nakamura graphs with $I=0$ have no $I$-structure. These graphs were counted using matrix models
in Section \ref{sec:hmm}.

\begin{figure}[h]
\begin{center}
\includegraphics[width=0.10\textwidth]{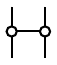}
\caption{The only possible $I$-structure for $I=1$.}
\label{fig:IstrI1}
\end{center}
\end{figure} 
Now consider the graphs where $I=1$, which have exactly one internal edge. From the definition of the reduced $S_d$ tuple 
description, the zeroes of an $I=1$ graph must be described by a pair of permutations $\tau_1$ and $\tau_2$, and for each such pair of permutations
there exists a unique $j\in\{1,2,\ldots d\}$ such that 
\bea 
&& \tau_1 (j) \ne j \cr 
&& \tau_2 (j) \ne j.
\eea
In other words, $j$ belongs to the moved-point sets of both $\tau_1$ and $\tau_2$. The associated $I$-structure consists of two columns
and a single $I$-row with two vertices. This is given in Figure \ref{fig:IstrI1}.

\begin{figure}[h]
\begin{center}
\includegraphics[width=0.50\textwidth]{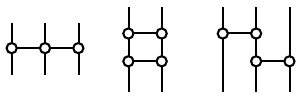}
\caption{The $I$-structures for $I=2$.}
\label{fig:IstrI2}
\end{center}
\end{figure} 
In the case that $I=2$, there are three distinct $I$-structures, as drawn in Figure \ref{fig:IstrI2}.
The first $I$-structure has three columns and one $I$-row with three vertices. This corresponds to tuples in which
there is a single integer $j\in\{1,2,\ldots, d\}$ that is permuted by all three permutations $\tau_1, \tau_2, \tau_3$, and
no other integer in the set $\{1,2,\ldots, d\}$ is permuted by any two of the $\tau_i$.
The second $I$-structure has two columns and two identical $I$-rows, each with two vertices. This structure corresponds to graphs for which
there are exactly two integers $j_1, j_2 \in \{1,2,\ldots, d\}$ that are mutually permuted by the pair of permutations $\tau_1$ and $\tau_2$.
The third $I$-structure has three columns and two distinct $I$-rows with two vertices. This corresponds to a triple  $\tau_1, \tau_2, \tau_3$,
with the property that there is some pair  $j_1, j_2 \in \{1,2, \ldots, d\} $ such that 
\bea 
\tau_1 ( j_1 ) \ne j_1,&  \qquad \tau_2 ( j_1 ) \ne j_1, \qquad &\tau_3 ( j_1 ) = j_1 \cr 
\tau_1 ( j_2 ) = j_2,& \qquad \tau_2 ( j_2 ) \ne j_2, \qquad &\tau_3 ( j_2 ) \ne j_2 
\eea

\begin{figure}[h]
\begin{center}
\includegraphics[width=0.7\textwidth]{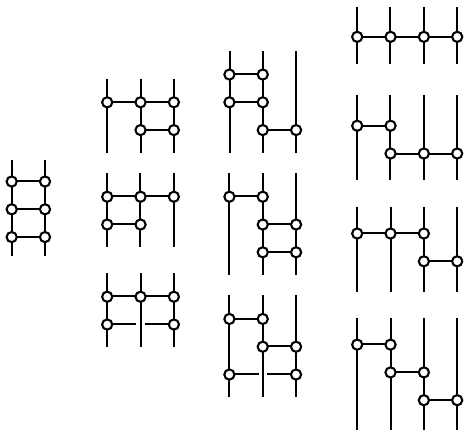}
\caption{The $I$-structures for $I=3$.}
\label{fig:IstrI3}
\end{center}
\end{figure} 
For $I=3$, there are eleven $I$-structures that can be drawn that correspond to tuples in the reduced $S_d$ description.
These are shown in Figure  \ref{fig:IstrI3}.

\subsection{An algorithm utilising \texorpdfstring{$I$}{I}-structures}

A Nakamura graph corresponds to a slide-equivalence class of permutation tuples.
Within each slide-equivalence class, there is a Hurwitz class of reduced $S_d$ tuples, 
which is an equivalence class under $S_d$ conjugation of permutation tuples 
$(\sigma_+, \tau_1, \ldots, \tau_m, \sigma_-)$.
If we consider just Nakamura graphs with a single incoming pole, then the permutation $\sigma_+$ is a $d$-cycle, and we
can use the $S_d$ symmetry to choose a representative element of the Hurwitz class with $\sigma_+=(1,2,\ldots, d)$.
The elements of the Hurwitz class with $\sigma_+=(1,2,\ldots, d)$ are permutation tuples conjugate to each other by elements in 
\bea \text{Aut}(\sigma_+) = \cor{(1,2,\ldots, d)} = Z_d. \eea
This means that a Nakamura graph corresponds to an equivalence class of reduced tuples $(\tau_1,\ldots,\tau_m)$ under the equivalence
\bea
(\tau_1, \ldots, \tau_m) \sim (\gamma^{-1} \tau_1 \gamma, \ldots, \gamma^{-1} \tau_m \gamma), \label{eq:Zdconj}
\eea
for $\gamma\in \cor{(1,2,\ldots, d)}=\mathbb{Z}_d$.
Each Nakamura graph has an associated class structure
$\cT_i = [\tau_i]$, $i=1,\ldots, m$, and an associated $I$-structure, describing which integers in the set
$\{1,2,\ldots, d\}$ are permuted by more than one permutation $\tau_i$.

There is an efficient algorithm that counts Nakamura graphs by using $I$-structures.
As in the original algorithm outlined above, the $I$-structures algorithm starts 
by finding all the unreduced and reduced class tuples.
For each reduced class tuple $(\cT_1, \ldots, \cT_m)$, the algorithm finds all possible $I$-structures that are consistent with this class tuple. 
Each $I$-structure must have one edge connecting columns $i-1$ and $i$ for each cycle in $\tau_i\in\cT_i$, where $i=2,\ldots,m$.
Also, there must be no more edges connecting each column $i=1,2,\ldots m$ in an $I$-structure than the 
total number of labels permuted by any $\tau_i\in\cT_i$. 

The algorithm considers each reduced class tuple and $I$-structure in turn.
All Nakamura graphs with this reduced class tuple have the same values of $\Delta$ and $I$, and all
Nakamura graphs with this chosen $I$-structure have the same value of $I$, and so all graphs with this
$I$-structure and class tuple have the same degree $d$, where
\bea
d = \Delta + 2l - I. \label{eq:degree}
\eea
Let $\Omega_{\cI, \cT}$ be the set of tuples $(\tau_1,\ldots,\tau_m)$ with a given $I$-structure $\cI$ and class structure
$(\cT_1, \ldots, \cT_m)$. The Nakamura graphs with the specified $I$-structure and class structure are
the equivalence classes of this set under the $\mathbb{Z}_d$ conjugation action (\ref{eq:Zdconj}).
However, the set $\Omega_{\cI, \cT}$ can be very large in general, so it is computationally very expensive to split
this set into $\mathbb{Z}_d$ conjugacy classes directly. 
One way of circumventing this difficulty is to break the problem into stages: we first split $\Omega_{\cI, \cT}$ into 
conjugacy classes under the equivalence relation
\bea
(\tau_1,\tau_2, \ldots, \tau_m) \sim (\alpha^{-1}\tau_1\alpha, \alpha^{-1}\tau_2\alpha, \ldots, \alpha^{-1} \tau_m \alpha),
\eea
where $\alpha\in S_d$.
Once we have found the $S_d$-equivalence classes of $\Omega_{\cI, \cT}$, we can act on the elements of each $S_d$-class 
individually with the group $\mathbb{Z}_d$ by conjugation and hence find the $Z_d$-equivalence classes of
$\Omega_{\cI, \cT}$, which are the distinct Nakamura graphs.
Also, rather than directly constructing the very large set $\Omega_{\cI, \cT}$ and then splitting it into 
$S_d$ equivalence classes, it is more efficient to construct these equivalence classes directly by finding a representative
element of each class.

We find the representative elements of the $S_d$-classes by using the $I$-structure and breaking the $S_d$ symmetry.
Let $k$ be the number of rows in the $I$-structure $\cI$, where $k\in\{0,1,\ldots, d\}$. 
For any tuple $(\tau_1,\ldots, \tau_m)\in \Omega_{\cI, \cT}$, there are exactly $k$ integers in $\{1,2, \ldots, d\}$
that are permuted by more than one $\tau_i$. These integers correspond to the internal edges of the Nakamura graph.
By adding the length of the cycles in the class $\cT_i$ for some $i\in\{1,2,\ldots,m\}$ and subtracting the
number of vertices in the $i$th column of the $I$-structure, we have the number of integers $e_i$ that are 
permuted by only the permutation $\tau_i$ within the tuple $(\tau_1, \ldots, \tau_m)$.
These integers correspond to the external edges of the graph.
Consider the set of `canonically-labelled' $\tau_i$-tuples $\tilde{\Omega}_{\cI, \cT}\subset \Omega_{\cI, \cT}$ which consists of those tuples in which the permuted integers $1$ to $k$ correspond to the rows of the $I$-structure, the integers $k+1, \ldots, k+e_1$ are permuted only by $\tau_1$, the labels $k+e_1+1, \ldots, k+e_1+e_2$ are permuted only by $\tau_2$, and so on. 
Each $S_d$-equivalence class of $\Omega_{\cI, \cT}$ contains at least one such canonically labelled $\tau_i$-tuple. 
A pair of canonically-labelled tuples are in the same $S_d$ equivalence class if and only if they are conjugate to 
each other by an element of the group $S_k\times S_{e_1}\times\ldots S_{e_m}$. This means that the orbits of the canonically-labelled tuples 
under the action by conjugation of the group $S_k\times S_{e_1}\times\ldots S_{e_m}$ are in direct correspondence with 
the equivalence classes of $\Omega_{\cI, \cT}$ under conjugation by $S_d$. 
As the set $\tilde{\Omega}_{\cI, \cT}$ is usually much smaller than $\Omega_{\cI, \cT}$, it is relatively cheap
computationally to construct the set of canonically-labelled tuples, find their orbits under $S_k\times S_{e_1}\times\ldots S_{e_m}$,
and choose a representative element from each orbit.
In this way, we can construct a set of representative elements of the $S_d$ classes of $\Omega_{\cI, \cT}$.

Consider each $S_d$-equivalence class of $\Omega_{\cI, \cT}$ in turn, specified by a representative $\tau$-tuple $(\tau_1, \ldots, \tau_m)$.
All the elements of this $S_d$ equivalence class are of the form $(\alpha^{-1} \tau_1 \alpha, \ldots, \alpha^{-1}\tau_m \alpha)$, 
where $\alpha \in S_d$.
Let $\text{Aut}(\tau)$ be the automorphism group of the representative $\tau$-tuple $(\tau_1, \ldots, \tau_m)$; that is, the set of 
elements $\gamma\in S_d$ that satisfy $\gamma^{-1}\tau_i\gamma = \tau_i$ for all $i=1,2,\ldots, m$.
If two permutations $\alpha, \tilde{\alpha}$ satisfy $\tilde{\alpha} = \gamma \alpha$ for some $\gamma\in  \text{Aut}(\tau)$,
then 
\bea
(\alpha^{-1} \tau_1 \alpha, \ldots, \alpha^{-1}\tau_m \alpha) = ({\tilde \alpha}^{-1} \tau_1 \tilde{\alpha}, \ldots, \tilde{\alpha}^{-1}\tau_m \tilde{\alpha}).
\eea
We can therefore see that each {\it right coset} $\text{Aut}(\tau)\alpha \in \text{Aut}(\tau)\backslash S_d$ specifies a unique element in the 
$S_d$ equivalence class of the $\tau$-tuple.

We wish to split this $S_d$ equivalence class into $\mathbb{Z}_d$ equivalence classes. A pair of elements of the $S_d$ 
equivalence class $(\alpha^{-1}\tau_1\alpha,\ldots,\alpha^{-1}\tau_m\alpha)$ and $(\tilde{\alpha}^{-1}\tau_i\tilde{\alpha},\ldots, \tilde{\alpha}^{-1}\tau_m\tilde{\alpha})$ are in the same $\mathbb{Z}_d$ equivalence class if and only if
\bea
(\tilde{\alpha}^{-1}\tau_i\tilde{\alpha},\ldots, \tilde{\alpha}^{-1}\tau_m\tilde{\alpha}) 
= 
(z^{-1}\alpha^{-1} \tau_1 \alpha z, \ldots, z^{-1}\alpha^{-1}\tau_m \alpha z)
\eea
for some $z \in \mathbb{Z}_d = \cor{(1,2,\ldots,d)}$. This means that two right cosets $\text{Aut}(\tau)\alpha$
and $\text{Aut}(\tau)\tilde{\alpha}$ are in the same $\mathbb{Z}_d$-equivalence class if
$\text{Aut}(\tau)\tilde{\alpha} = (\text{Aut}(\tau)\alpha)z$ for some $z \in \mathbb{Z}_d$.
We deduce that the {\it double cosets} 
\bea 
\text{Aut}(\tau)\alpha\mathbb{Z}_d \quad \in \quad \text{Aut}(\tau)\backslash S_d / \mathbb{Z}_d
\eea
parametrise the $\mathbb{Z}_d$-equivalence classes of a given $S_d$-equivalence class of $\Omega_{\cI, \cT}$,
and so give the Nakamura graphs associated to a given $S_d$-equivalence class of $\Omega_{\cI, \cT}$.

We can read off the size of the automorphism group of each graph by looking at the size of its associated double coset.
The product group $\text{Aut}(\tau)\times \mathbb{Z}_d$ acts on the elements in $S_d$ by left and right multiplication. The orbits of this action are the double cosets $\text{Aut}(\tau)\backslash S_d / \mathbb{Z}_d$. The stabiliser group of an element $\alpha\in S_d$ under this
action consists of the pairs of elements $(\gamma,z)$ which satisfy $\gamma \alpha z = \alpha$, or equivalently $\alpha^{-1}\gamma\alpha =  z^{-1}$.
As $\gamma$ and $z$ can be any elements of the groups $\text{Aut}(\tau)$ and $\mathbb{Z}_d$, the stabiliser of $\alpha$ is 
precisely the intersection of the groups $\alpha^{-1}\text{Aut}(\tau)\alpha$ and $\mathbb{Z}_d$. These are exactly the elements which fix 
under conjugation every element in the tuple $(\sigma_+, \alpha^{-1}\tau_1\alpha, \ldots, \alpha^{-1}\tau_m\alpha, \sigma_-)$,
and so the stabiliser of $\alpha$ is the automorphism group of the graph.
By the orbit-stabiliser theorem, we therefore deduce that the size of the automorphism group of a Nakamura graph 
given by the double coset $\text{Aut}(\tau)\alpha\mathbb{Z}_d$ is
\bea
\frac{d|A|}{|\text{Aut}(\tau)\alpha\mathbb{Z}_d|}.
\eea
The software GAP can efficiently count double cosets and find their representative elements and sizes. The algorithm we have devised
is therefore able to quickly find all the Nakamura graphs that arise from a given representative $\tau$-tuple
in the $S_d$-equivalence classes of  $\Omega_{\cI, \cT}$, and to read off their automorphism group sizes.

As an example of this procedure, we consider the reduced class tuple 
$(\cT_1, \cT_2, \cT_3)=([2,2], [3], [2])$ with $|\chi|=5$. This class tuple contains only one cycle
with cycle size greater than 2, so its branching number is $\Delta=1$.
From the relation
\bea
2|\chi| -d = \Delta + I,
\eea
we know that the degree and the number of internal edges are related by $d+I=9$. 
Permutations in the class $\cT_1$ permute four integers, so the degree is bounded 
from below by 4. There are three classes in this reduced class tuple, so there are
at least two internal edges. This means that the number of internal edges $I$ lies
in the range $\{2,3,4,5\}$.

\begin{figure}[h]
\begin{center}
\includegraphics[width=0.25\textwidth]{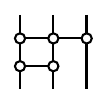}
\caption{An example of an $I$-structure of $(\cT_1, \cT_2, \cT_3)=([2,2], [3], [2])$ with $d_{max}=10$.}
\label{fig:cI}
\end{center}
\end{figure} 
One of the $I$-structures found by the algorithm is given in Figure \ref{fig:cI}.
This structure has $I=3$ internal edges, and degree $d=6$.
Let $\Omega_{\cI, \cT}$ be the set of tuples corresponding to this $I$-structure and reduced class structure.
This $I$-structure has two rows, so there are $k=2$ integers corresponding to internal edges
in each tuple. The first column has two vertices, and corresponds to the class $\cT_1=[2,2]$ of permutations 
which permute four integers. This means that there are $e_1=2$ integers permuted by the first permutation
in each tuple which correspond to external edges. Similarly, there are $e_2=1$ integers permuted only 
by the permutation $\tau_2$ within each tuple and $e_3=1$ integers permuted by the permutation $\tau_3$.

To find the $S_d$-equivalence classes of $\Omega_{\cI, \cT}$, we first find the `canonically-labelled' tuples 
$(\tau_1, \tau_2, \tau_3)$ in which $\tau_1$ permutes the integers $\{1,2,3,4\}$, $\tau_2$ permutes $\{1,2,5\}$, and $\tau_3$ 
permutes $\{1, 6\}$. There are six such elements, and the set of canonically-labelled tuples is
\bea
\tilde{\Omega}_{\cI, \cT} &=& \{ (1,2)(3,4), (1,3)(2,4), (1,4)(2,3) \}\times \{ (1,2,5), (1,5,2)\} \times \{ (1,6) \}.
\eea
 Next, we consider the orbits in $\Omega_{\cI, \cT}$ generated by this set under the action of the group
$S_k\times S_{e_1}\times S_{e_2}\times S_{e_3} = \cor{(1,2), (3,4)}$.
Note that the set $\tilde{\Omega}_{\cI, \cT}$ is not closed under this group action.
The tuples $( (1,3)(2,4), (1,2,5), (1,6) )$ and $( (1,4)(2,3), (1,2,5), (1,6) )$ are conjugate, as are the tuples
$( (1,3)(2,4), (1,5,2), (1,6) )$ and $( (1,4)(2,3), (1,5,2), (1,6) )$, and so a set of representatives for the orbits
of the canonically-labelled tuples $(\tau_1, \tau_2, \tau_3)$ is
\bea
( (1,2)(3,4), (1,2,5), (1,6) ), \nn \\
( (1,2)(3,4), (1,5,2), (1,6) ), \nn \\
( (1,3)(2,4), (1,2,5), (1,6) ), \nn \\
( (1,3)(2,4), (1,5,2), (1,6) ).
\eea
These are representative elements of the $S_d$-equivalence classes of $\Omega_{\cI, \cT}$.

For each representative tuple, the Nakamura graphs are given by the double cosets 
$\text{Aut}(\tau)\backslash S_d / \mathbb{Z}_d$.
The representative tuple $(\tau_1, \tau_2, \tau_3) = ( (1,2)(3,4), (1,2,5), (1,6) )$ has the automorphism
group $\text{Aut}(\tau) = \cor{(3,4)}$, and so the Nakamura graphs are the double cosets $\cor{(3,4)}\backslash S_6 / \cor{(1,2,\ldots, 6)}$.
There are 60 distinct double cosets, all consisting of 12 elements, and so there are 60 Nakamura graphs in this $S_d$-class.
All these graphs have trivial automorphism group.
For the representative  tuple $(\tau_1, \tau_2, \tau_3) = ( (1,3)(2,4), (1,2,5), (1,6) )$, the automorphism
group $\text{Aut}(\tau)$ is trivial, and so the double cosets are $\{()\}\backslash S_6/ \cor{(1,2,\ldots, 6)}.$
There are 120 distinct double cosets in this case, and so there are 120 Nakamura
graphs in this $S_d$-equivalence class.

\subsection{GAP results for \texorpdfstring{$|\chi|=7$}{|chi|=7}}
The $I$-structure counting algorithm produces a complete catalogue of the Nakamura graphs for any given genus. 
In Appendix \ref{sec:apptables}, we have presented the output for the graphs with graph Euler characteristic $|\chi|=7$,
catalogued by genus $g$, number of poles $n$, the dimensions of their associated cells in moduli space,
and their automorphism groups. This extends the data found in \cite{Nak}.

We can perform a non-trivial check on the validity of this approach and of the cell-decomposition
of moduli space by comparing these tables with the orbifold Euler characteristic of moduli space.
Harer and Zagier \cite{HZ} give the following formulae for the orbifold Euler characteristic of $\cM_{g,n}$:
\bea
\chi(0,n) &=& \frac{(-)^{n-1}}{(n-1)(n-2)}, \qquad n\geq 3, \nn\\ 
\chi(1,n) &=& \frac{(-)^n}{12}, \qquad n\geq 2, \nn \\
\chi(g,n) &=& \frac{(-)^{2g}}{2g}\binom{2g+n-3 }{ n-1}B_{2g} \qquad g\geq 2, n\geq 0, \label{hz}
\eea
where $B_{2g}$ is a Bernoulli number. 
(The formulae given in \cite{HZ} are a factor of $(n-1)!$ larger than the formulae given here, since
we have allowed the outgoing poles of the graphs to be interchanged by automorphisms.)

Using these tables, and the defining formula for an orbifold Euler characteristic
\bea
\chi(g, n) = \sum_{\cG}(-1)^\text{dim}\frac{1}{|\text{Aut}(\cG )|}
\eea
we find 
\bea \chi(0,9) &=& \frac{1}{56}, \nn \\
\chi(1,7)&=&-\frac{1}{12}, \nn \\
\chi(2,7)&=&\frac{1}{8}, \nn \\
\chi(3,3) &=& -\frac{5}{84}. \nn
\eea
This is consistent with the formulae \eref{hz} from Harer and Zagier.

\section{Summary and future directions} 

Nakamura \cite{Nak} gave a description of light-cone string diagrams in terms of embedded graphs on the worldsheet, which are constructed from the Giddings-Wolpert differential on the worldsheet. 
  He used it to describe a cell decomposition of the space of GW-differentials. These cells can be quotiented 
by the automorphism groups of  the graphs to obtain cells in $\cM_{g,n}$. This allowed a 
  computation of  the orbifold Euler characteristics  of $\cM_{g,n}$ for small values of $g$ and $n$. We have developed connections between Nakamura graph combinatorics, branched covers and permutation tuples. 
By considering the light-cone diagrams with a single incoming string, we used known results on hermitian matrix model correlators to give analytic results for the contribution of the top-dimensional cells in the LC decomposition. This could be generalised to
cases with two or more incoming strings by using \cite{shakmor} and generalisations thereof. 
Beyond the top-dimensional cells, we related the contributions to the orbifold Euler characteristic from lower-dimensional cells with $\Delta>0$ and $I=0$ to analytic expressions in complex matrix models. 

As observed in \cite{Nak}, the numbers of cells in the LC cell decomposition 
for given $g$ and $n$ are smaller than the corresponding number in the KP 
cell decomposition. This is because the Nakamura graphs, which corresponds to the cells, are embedded graphs, but with restrictions related to the fact that the edges are real trajectories of the GW differential.
The fact that there is a well-defined global time coordinate imposes restrictions on the connectivity of 
embedded graphs which can be Nakamura graphs. These restrictions are detailed in the language
of permutations in Section \ref{sec:triplestocells}.
This suggests that it would  be worthwhile to revisit mathematical 
questions on the topology of $ \cM_{ g , n }$ using the LC cell decomposition. The computation of 
all the homology groups is still an open question. For a recent paper, see for example \cite{murri}, and for
associated discussion \cite{overflow-betti}. From a physics perspective, an immediate goal would be 
to use the improved understanding of the LC cell decomposition in the computation of string amplitudes 
in the light-cone gauge, either in the first quantised or second quantised string field formalism. 

The LC cell decomposition gives precise information about the topology of $\cM_{g,n}$. 
The codimension of a cell is $2 \Delta + I$, with $\Delta$ increasing 
when the zeroes of the GW differential have higher order. The parameter $I$ is the number of internal edges 
of the Nakamura graph, connecting the zeroes of the differential.
An improved understanding of the structure of the lower dimensional cells in the LC decomposition 
can be expected to shed light on the issue of ``contact terms'' in the light-cone approach to 
string amplitudes. It is believed that second quantised bosonic light-cone string field theory requires no contact terms,
but superstrings require contact terms (see for example the review \cite{spradvol03}). 
Contact terms related to higher order ramification points have been discussed in \cite{motldijkgraaf}
in connection with the matrix string theory of Dijkgraaf, Verlinde, and Verlinde \cite{dvv}. 
As pointed out in  \cite{witten1307} there is no direct superstring analogue of the bosonic worldsheet moduli space, but rather superstring theory requires integration over an appropriate cycle in a product $ \cM_{ L } \times \cM_{ R }$ of moduli spaces, with $\cM_{L}$ and $\cM_{R}$ closely related to $\cM_{g,n}$. 
It would be interesting to investigate how an improved  understanding of the combinatorics of cell decompositions   in $ \cM_{ g , n }$ 
can lead to cell decompositions for  the  integration cycles appropriate for superstring theory amplitudes. 

A very interesting problem is to give a precise description of the cell decomposition of 
$\cM_{g,n}$ arising from the light-cone approach. We know that there is a cell for every Nakamura graph.
The Nakamura graph has parameters which are interaction times and strip widths. These are related to the 
 more traditional parametrisation in terms of times, internal string momenta (widths) and twist angles \cite{mandelstam73,mandelstam85,dhokPhong,ishmur1307}. The automorphism group of the Nakamura 
graphs should have a natural action on the strip widths and time parameters, which would allow the 
space  of these parameters associated to a given graph to be quotiented out by the group.  Clarifying this in generality (i.e. for any graph at any genus $g$,  any number of punctures $n>1 $ and for any choice of external momenta) will be a very useful step in better understanding  the geometry of the  light-cone cell decomposition. It 
would  solve the problem (discussed in \cite{GiddWol,dhokPhong}) of giving the precise restrictions  
on the light-cone diagram parameters to ensure that every Riemann surface appears precisely once and
should lead to  progress in the computation of string amplitudes in the light-cone. The results of  the present paper suggest that  the  general permutation group descriptions of  Nakamura graphs will be the right  set-up to approach this question.   We hope to return to this problem in the near future.

Belyi maps, and the related equivalence classes of permutation triples,
have played an important role in this paper. A general Nakamura graph is related to permutation 
triples in $S_{4d+2I}$ or $S_{2d+2I}$, albeit only those equivalence classes of triples subject to intricate causality conditions. 
It is known that Belyi maps have deep connections to number theory and as such form an active subject of 
research in mathematics \cite{Grothendieck,schneps,LandoZvon}. Investigation of the link between light cone cell decompositions 
of $ \cM_{g,n}$ and Belyi maps can lead to a new interplay between string theory and number theory. One of the themes 
of interest in the number theory context is that Belyi maps form complete orbits of the absolute Galois group. It is also 
known that certain restricted classes of Belyi maps, e.g. those related to tree-like dessins, form complete orbits \cite{schneps}.
Is the same true of the restricted classes related to Nakamura graphs? 
Belyi maps came up again in the $S_d$ description of Nakamura graphs, when we specialised 
to top-dimensional cells of the LC cell decompositions and related the counting of the cells to Hermitian matrix models. 
This link between Belyi maps and Hermitian matrix models has been investigated as an avenue towards 
a topological string description of the Hermitian matrix model \cite{KRBelyi1002,Gopak,GopakPius,Robert}, 
as the simplest model of gauge-string duality.  
It is also an example of the ubiquity of combinatoric low dimensional topological field theories, based on 
Dijkgraaf-Witten models with permutation groups, in gauge theoretic correlators \cite{dMKRam2010,dRW1209,quivcal, tensmod, holofuzzy}. 
The present work extends these topological field theory and topological string structures to the fundamentals of string amplitudes 
and moduli space $\cM_{g,n}$. It is likely that future developments will see a deeper interplay between simple 
models of gauge-string duality, combinatoric topological field theories and traditional string amplitude computations. 

Nakamura graphs, with the construction of general closed string worldsheets at any point in $\cM_{g,n}$ in terms of 
flat strips glued together, are central to the metastring, a new foundational approach to the geometry of string theory and spacetime 
being developed in \cite{FLM1307}. This relationship is developed in more detail in the recent paper \cite{meta}.

\vskip2cm 

\begin{centerline}
{ \bf Acknowledgements} 
\end{centerline} 

\vskip1cm 

We thank Nathan Berkovits,  Robert de Mello Koch, Nick Evans, Edward Hughes, Rodolfo Russo, and Gabriele Travaglini for discussions. 
LF would like to thank R.G. Leigh and D. Minic for many stringy interactions.
DG and SR would like to thank Perimeter Institute for hospitality during May-June 2013, 
and DG would like to thank the Southampton High Energy Physics group (SHEP) for their hospitality and for all the interesting discussions during the summer of 2014.
SR is supported by STFC consolidated grant ST/L000415/1
"String Theory, Gauge Theory \& Duality."
Research at Perimeter Institute for Theoretical Physics is supported in part by the Government of Canada through NSERC and by the Province of Ontario through MRI.

\newpage
\appendix
\section{Tables of Nakamura graphs with $|\chi|=7$}\label{sec:apptables}

\begin{table}[h!]
\caption{$(g,n)$=(0,9)}
 \label{tab:istrfirst}
 \begin{center}
  \begin{tabular}{|c|c|c|c|c|c|c|}
    \hline
    dimension & 12 & 11 & 10 & 9 & 8 & 7 \\ \hline     
$\begin{array}{c}\mbox{Graphs} \\ ([\mbox{Aut}] \times\mbox{Number)}\end{array}$
&
$\begin{array}{l} \gau{1}{28} \\ \gau{2}{5} \\ \gau{7}{1} \end{array}$
&
$\gau{1}{297}$
&
$\begin{array}{l} \gau{1}{1324} \\ \gau{2}{25} \end{array}$
&
$\gau{1}{3675}$
& 
$\begin{array}{l} \gau{1}{6795} \\ \gau{2}{52} \\ \gau {4}{1} \end{array}$
&
$\gau{1}{8892}$
\\ \hline \end{tabular} \hspace{5mm}

\vspace{5mm}

\hspace{5mm}\begin{tabular}{|c|c|c|c|c|c|c| }
\hline
6&5&4&3&2&1&0 \\ \hline
$\begin{array}{l} \gau{1}{8169} \\ \gau{2}{57} \end{array}$
&    
$\gau{1}{5250}$
&
$\begin{array}{l}  \gau{1}{2226} \\ \gau{2}{29} \\  \gau{4}{2} \end{array}$ 
&
$\gau{1}{595}$
&
$\begin{array}{l}\gau{1}{85} \\ \gau{2}{6} \end{array}$
&
$\gau{1}{6}$
&
$\gau{8}{1}$
\\ \hline \end{tabular}
\end{center}
\end{table}

\begin{table}[h!]
\caption{$(g,n)$=(1,7) }
\begin{center}
\begin{tabular}{|c|c|c|c|c|c|c|c|}
    \hline
    dimension & 14 & 13 & 12 & 11 & 10 & 9 & 8 \\ \hline     
$\begin{array}{c}\mbox{Graphs} \\ \mbox{([Aut]}\times\mbox{\#)}\end{array}$
&
$\begin{array}{l} \gau{1}{838} \\ \gau{2}{40} \end{array}$
&
$\gau{1}{9702}$
&
$\begin{array}{l} \gau{1}{51870} \\ \gau{2}{210} \end{array}$
&
$\gau{1}{174090}$
&
$\begin{array}{l} \gau{1}{404059} \\ \gau{2}{471}  \\ \gau{3}{1}  \\ \gau{4}{2} \\ \gau{6}{1} \end{array}$
&
$\gau{1}{680960}$
&
$\begin{array}{l} \gau{1}{843976} \\ \gau{2}{574} \end{array}$
\\ \hline \end{tabular}

\vspace{5mm}

\hspace{5mm}\begin{tabular}{|c|c|c|c|c|c|c| }
\hline
7&6&5&4&3&2&1 \\ \hline
$\begin{array}{l} \gau{1}{766000} \\ \gau{3}{5} \end{array}$
&
$\begin{array}{l} \gau{1}{497046} \\ \gau{2}{378} \\ \gau{4}{4} \end{array}$
&
$\gau{1}{222057}$
&
$\begin{array}{l} \gau{1}{64087} \\ \gau{2}{124} \\ \gau{3}{5} \\ \gau{6}{2} \end{array}$
&
$\gau{1}{10820}$
&
$\begin{array}{l} \gau{1}{863} \\ \gau{2}{15} \\ \gau{4}{1} \end{array}$
&
$\begin{array}{l} \gau{1}{18} \\ \gau{3}{2}   \end{array}$
\\ \hline \end{tabular}
\end{center}
\end{table}

\begin{table}[h!]
\caption{$(g,n)$=(2,5)}
 \begin{center}
  \begin{tabular}{|c|c|c|c|c|c|}
    \hline
    dimension & 16 & 15 & 14 & 13 & 12  \\ \hline     
$\begin{array}{c}\mbox{Graphs} \\ \mbox{([Aut]}\times\mbox{\#)}\end{array}$
&
$\begin{array}{l} \gau{1}{4680} \\ \gau{2}{78} \end{array}$
&
$\gau{1}{59598}$
&
$\begin{array}{l} \gau{1}{359771} \\ \gau{2}{485} \end{array}$
&
$\gau{1}{1374975}$
&
$\begin{array}{l} \gau{1}{3688668} \\ \gau{2}{1322}  \\ \gau{3}{9}  \\ \gau{4}{2} \end{array}$
\\ \hline \end{tabular} \hspace{5mm}

\vspace{5mm}

\begin{tabular}{|c|c|c|c|c|c|}
\hline
11&10&9&8&7&6 \\ \hline
$\gau{1}{7291788}$
&
$\begin{array}{l} \gau{1}{10799810} \\ \gau{2}{1995} \end{array}$
&
$\begin{array}{l} \gau{1}{11954262} \\ \gau{3}{30} \end{array}$
&
$\begin{array}{l} \gau{1}{9708622} \\ \gau{2}{1671} \\ \gau{4}{5} \end{array}$
&
$\gau{1}{5611630}$
&
$\begin{array}{l} \gau{1}{2204212} \\ \gau{2}{695} \\ \gau{3}{36} \end{array}$
\\ \hline \end{tabular}

\vspace{5mm}

\hspace{5mm}\begin{tabular}{|c|c|c|c|}
\hline
5&4&3&2 \\ \hline
$\gau{1}{548779}$
&
$\begin{array}{l} \gau{1}{76822} \\ \gau{2}{101} \\ \gau{4}{3} \\ \gau{8}{1} \end{array}$
&
$\begin{array}{l} \gau{1}{4814} \\ \gau{3}{12} \end{array}$
&
$\gau{1}{84}$
\\ \hline \end{tabular}

 \end{center}
\end{table}

\begin{table}[h!]
\caption{$(g,n)$=(3,3)}
 \label{tab:istrlast}
 \begin{center}
  \begin{tabular}{|c|c|c|c|c|c|}
    \hline
    dimension & 18 & 17 & 16 & 15 & 14 \\ \hline     
$\begin{array}{c}\mbox{Graphs} \\ \mbox{([Aut]}\times\mbox{\#)}\end{array}$
&
$\begin{array}{l} \gau{1}{4013} \\ \gau{2}{63} \\ \gau{7}{2} \\ \gau{14}{1} \end{array}$
&
$\gau{1}{55143}$
&
$\begin{array}{l} \gau{1}{360892} \\ \gau{2}{421} \end{array}$
&
$\gau{1}{1502760}$
&
$\begin{array}{l} \gau{1}{4420204} \\ \gau{2}{1236}  \\ \gau{3}{7}  \\ \gau{4}{5} \\ \gau{6}{1} \end{array}$
\\ \hline \end{tabular} \hspace{5mm}

\vspace{5mm}

\begin{tabular}{|c|c|c|c|c|c|}
\hline
13&12&11&10&9&8 \\ \hline
$\gau{1}{9649120}$
&
$\begin{array}{l} \gau{1}{15910334} \\ \gau{2}{2031} \end{array}$
&
$\begin{array}{l} \gau{1}{19771176} \\ \gau{3}{25} \end{array}$
&
$\begin{array}{l} \gau{1}{18191095} \\ \gau{2}{1891} \\ \gau{4}{11} \end{array}$
&
$\gau{1}{12042490}$
&
$\begin{array}{l} \gau{1}{5502643} \\ \gau{2}{940} \\ \gau{3}{29} \\ \gau{6}{2} \end{array}$
\\ \hline \end{tabular}

\vspace{5mm}

\hspace{5mm}\begin{tabular}{|c|c|c|c|}
\hline
7&6&5&4 \\ \hline
$\gau{1}{1632983}$
&
$\begin{array}{l} \gau{1}{284718} \\ \gau{2}{203} \\ \gau{4}{4} \\ \gau{8}{2} \end{array}$
&
$\begin{array}{l} \gau{1}{24312} \\ \gau{3}{10} \end{array}$
&
$\begin{array}{l} \gau{1}{680} \\ \gau{2}{12} \end{array}$
\\ \hline \end{tabular}

\end{center}
\end{table}

\FloatBarrier

\end{document}